\documentclass[letterpaper,11pt]{article}
\pdfoutput=1
\usepackage{jheppub}
\pdfoutput=1
\usepackage{epstopdf}
\usepackage{slashed}
\usepackage{graphicx} 
\usepackage{subfig} 
\usepackage{soul}
\usepackage{amsmath}
\usepackage{multirow}
\usepackage{tablefootnote}
\usepackage{comment}
\usepackage{color}
\usepackage[utf8]{inputenc}

\newcommand{\beq}{\begin{equation}}
\newcommand{\eeq}{\end{equation}}
\newcommand{\bea}{\begin{eqnarray}}
\newcommand{\eea}{\end{eqnarray}}

\newcommand{\be}{\begin{equation}}
\newcommand{\ee}{\end{equation}}
\newcommand{\beal}{\begin{align}}
\newcommand{\eeal}{\end{align}}

\def\non{\nonumber}

\def\vep{\varepsilon}

\def\tr{\mathrm{Tr}}
\def\ordr{\mathcal{O}}

\def\<{\langle}
\def\>{\rangle}

\def\mt{{\mathcal{T}}}
\def\LI{{\mathrm{LI}}}
\def\nlsm{NLSM}

\newcommand*\DAl{\mathop{}\!\mathbin\Box}

\newcommand{\lsim}{\mathrel{\mathop{\kern 0pt \rlap
  {\raise.2ex\hbox{$<$}}}
  \lower.9ex\hbox{\kern-.190em $\sim$}}}
\newcommand{\gsim}{\mathrel{\mathop{\kern 0pt \rlap
  {\raise.2ex\hbox{$>$}}}
  \lower.9ex\hbox{\kern-.190em $\sim$}}}

\definecolor{pink}{RGB}{255,105,180}

\preprint{CERN-TH-2018-090}

\title{The Infrared Structure of Nambu-Goldstone Bosons}

\author[a,b,c]{Ian Low}
\author[a]{and Zhewei Yin}

\affiliation[a]{Department of Physics and Astronomy, Northwestern University, Evanston, IL 60208, USA}
\affiliation[b]{High Energy Physics Division, Argonne National Laboratory, Argonne, IL 60439, USA}
\affiliation[c]{Theoretical Physics Department, CERN, 1211 Geneva 23, Switzerland}

\emailAdd{ilow@anl.gov}
\emailAdd{zheweiyin2015@u.northwestern.edu}

\abstract{The construction of effective actions for Nambu-Goldstone bosons, and the  nonlinear sigma model, usually requires  a target coset space $G/H$. Recent progresses uncovered a new formulation using only IR data without reference to the broken group $G$ in the UV, by imposing the Adler's zero condition, which  can be seen to originate from the superselection rule in the space of degenerate vacua.   The IR construction imposes a nonlinear shift symmetry on the Lagrangian to enforce the correct single soft limit amid constraints of the unbroken group $H$. We present a systematic study on the consequence of the Adler's zero condition in correlation functions of nonlinear sigma models, by deriving the conserved current and the Ward identity associated with the nonlinear shift symmetry, and demonstrate how the old-fashioned current algebra emerges. The Ward identity leads to a new representation of on-shell amplitudes,  which amounts to bootstrapping the higher point amplitudes from lower point amplitudes and adding new vertices to satisfy the Adler's condition. The IR perspective allows one to extract Feynman rules for  the mysterious extended theory of biadjoint cubic scalars residing in the subleading single soft limit, which was first discovered using the Cachazo-He-Yuan representation of scattering amplitudes. In addition, we present the subleading triple soft theorem in the nonlinear sigma model and show that it is also controlled by  on-shell amplitudes of the same extended theory as in the subleading single soft limit.

}


\begin{document}
\maketitle
\flushbottom

\section{Introduction}

\begin{figure}[htbp]
\centering
\includegraphics[width=0.4\textwidth]{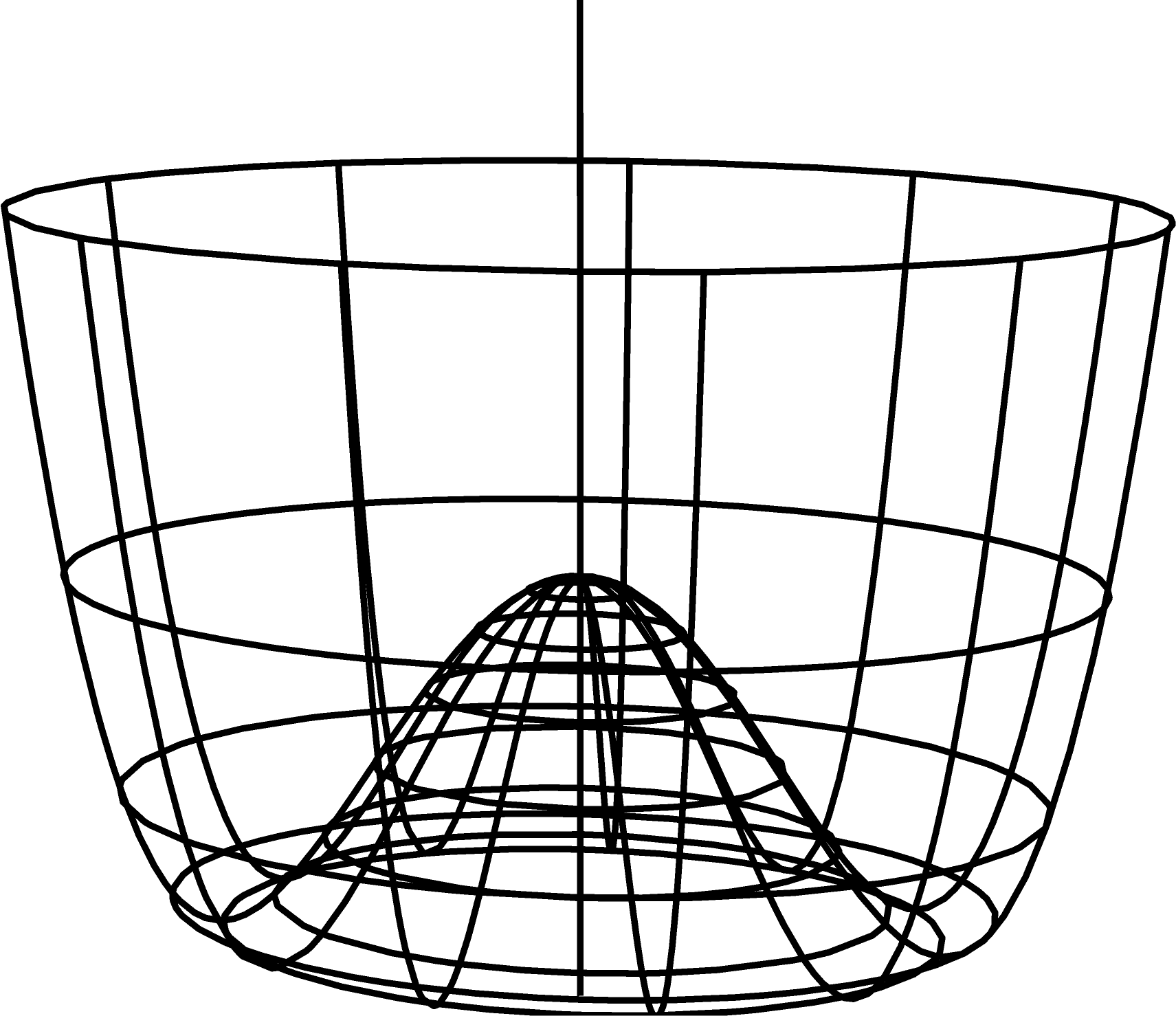}
\caption{\label{fighat} An example of a Mexican hat potential, with $\mathbf{\Phi}=(\Phi^1, \Phi^2)$.}
\end{figure}

Nambu-Goldstone bosons \cite{PhysRev.117.648,Goldstone:1961eq} are long-range excitations over degenerate vacua transforming non-trivially under continuous global symmetries. The discussion on Nambu-Goldstone bosons in textbooks on quantum field theories often starts with  the Goldstone's theorem  which states that, for a global symmetry group $G$, there is a massless spin-0 particle associated with each generator $X^a$ of $G$ such that the vacuum state is not annihilated,
\be
X^a |0\rangle\neq 0 \ ,
\ee
while those generators that do annihilate the vacuum,
\be
T^i |0\rangle = 0 \ ,
\ee
form a subgroup $H$ of $G$. In this language $H$ is commonly referred to as the unbroken group and $G$ the broken group. Moreover, $X^a$ is the broken generator residing in the coset space $G/H$. The classic example starts with a set of scalars, $\mathbf{\Phi}=(\Phi^1,\cdots, \Phi^n)$, furnishing a linear representation of $G$ with the  Mexican hat potential  shown in Fig. \ref{fighat}:
\be
V(\mathbf{\Phi}) = -\frac{\mu^2}2  |\mathbf{\Phi}|^2 + \frac{\lambda}4 |\mathbf{\Phi}|^4 \ .
\ee
The ground state consists of all configurations where the vacuum expectation value (VEV) of $\mathbf{\Phi}$ is non-vanishing, $\langle \mathbf{\Phi}\rangle = {\mu}/{\sqrt{\lambda}}\ \mathbf{n}$, where $\mathbf{n}\cdot \mathbf{n}=1$. Generators of the unbroken group $H$ consists of those that annihilate the VEV: $T^i\langle \mathbf{\Phi}\rangle =0$. The rest of the generators of $G$ are labelled as the broken generators $X^a$. The Nambu-Goldstone mode $\pi^a$ is the ``angular variable":
\be
\label{eq:phigoldstone}
\mathbf{\Phi}= \left(\frac{\mu}{\sqrt{\lambda}}+h(x)\right) e^{i X^a\pi^a(x)}\ \mathbf{n} \ .
\ee
After plugging back into the potential $V(\mathbf{\Phi})$ one finds $\pi^a(x)$ is indeed massless. In this  treatment  it is essential that one acquires knowledge of the broken group $G$ in advance. In particular, the form of Eq.~(\ref{eq:phigoldstone}) also makes it clear that the Nambu-Goldstone mode connects different degenerate vacua at the minima of the Mexican hat potential, which are related to one another by an ``angular rotation."

In the seminal works of Coleman, Callan, Wess and Zumino (CCWZ) \cite{Callan:1969sn,Coleman:1969sm} which gives the most general construction of effective Lagrangians for Nambu-Goldstone bosons, the fundamental object is  the Cartan-Maurer one-form
\be
-i\xi^{-1}\partial_\mu \xi = {\cal D}_\mu\pi^a X^a + E_\mu^i\, T^i \ , \quad \xi=\xi(\pi)= e^{i\pi^a X^a/f} \ ,
\ee
where $f$ is the decay constant of the Nambu-Goldstone boson. Under the action of $g\in G$, $\xi$ transforms as
\be
\label{eq:complicate}
g\ \xi(\pi) =  \xi(\pi^\prime)\ U(\pi, g)\ ,
\ee
where $U(\pi,g)$ is an element of $H$ and $\pi^{\prime\, a}=\pi^{\prime\, a}(\pi, g)$ is a highly nonlinear function of $g$ and the original $\pi^a$. On the other hand, the Goldstone covariant derivative ${\cal D}_\mu\pi^a$ and the gauge connection $E_\mu^i$ have simple transformation rules,
\be
{\cal D}_\mu\pi^a X^a \to U\,({\cal D}_\mu\pi^a X^a) U^{-1}  \ , \qquad
E_\mu^i T^i \to U \, (E_\mu^i T^i)\,  U^{-1} - i U \partial_\mu U^{-1}  \ ,
\ee
from which one proceeds to build the effective Lagrangian
\be
{\cal L}_{eff} = \frac{f^2}2 {\cal D}_\mu \pi^a {\cal D}^\mu \pi^a + \cdots \ ,
\ee
Terms neglected above carry four derivatives and more. Nambu-Goldstone bosons in this approach can be viewed as coordinates parameterizing the coset space $G/H$. It is abundantly clear that, in CCWZ, prior knowledge on a broken group $G$ and an unbroken group $H$ is mandatory. Therefore, the commonly subscribed viewpoint is as follows:  in the ultraviolet when the  energy scale $E\gg \Lambda = 4\pi f\sim 4\pi |\langle \mathbf{\Phi}\rangle|$,  the symmetry $G$ is realized linearly and fully restored. Below $\Lambda$, however, $G$ is spontaneously broken and only the unbroken group $H$ is linearly realized. In this regime, one often uses the language that $G$ is {\em nonlinearly} realized and attributes the  nonlinear nature of the Goldstone interactions to the  broken group $G$. This perspective suggests the nonlinear Goldstone interaction is dependent on  $G$ in the ultraviolet, as indicated in the CCWZ formalism.

The ``top-down" approach adopted by CCWZ is very  powerful, and has been widely employed in constructing effective actions of Nambu-Goldstone bosons and nonlinear sigma models (\nlsm). On the other hand, it is worth recalling that Nambu-Goldstone bosons owe their presence to the non-trivial vacua in the deep infrared, as they are  long wave-length degrees of freedom connecting the different degenerate vacua. As such, it seems paradoxical that their interaction would depend on details of the broken group $G$ in the ultraviolet. 

In fact, Goldstone interactions in the deep infrared were studied intensively in the context of pions in low-energy QCD, which are  Nambu-Goldstone bosons from spontaneously broken $SU(2)_L\times SU(2)_R$ chiral symmetry. This body of work is collectively known as ``soft pion theorems" \cite{Treiman:1986ep}, some of which turn out to be universal properties of Nambu-Goldstone bosons and insensitive to the detail of the symmetry breaking pattern. One particularly well-known theorem is the Adler's zero condition \cite{Adler:1964um}, which dictates that on-shell scattering amplitudes of pions must vanish as one external momentum is taken to be soft. The soft pion theorems were mostly derived using the current algebra technique, which becomes quite cumbersome beyond the leading order result \cite{Dashen:1969ez}. Nevertheless, the Adler's zero condition serves as the prime example that Nambu-Goldstone bosons behave universally in the deep infrared. 

Inspired by this observation, a new approach was proposed in Refs.~\cite{Low:2014nga,Low:2014oga} to use the Adler's zero condition as the {\em defining property} of Nambu-Goldstone bosons, by constructing an effective Lagrangian whose S-matrix elements satisfy the Adler's zero condition in the presence of a linearly realized symmetry group $H$. A similar approach was also demonstrated in Ref. \cite{Ellis:1970nn} with $H= SU(2)$, while in our treatment we consider a general $H$. Surprisingly, the entire Nambu-Goldstone interactions can be constructed this way using only the infrared data: once a linear representation of the unbroken group $H$ is chosen, under which the Goldstone boson transforms, the complete effective action can be obtained by imposing the Adler's zero condition on the S-matrix elements, without reference to the target coset $G/H$. The only undetermined parameter is the overall normalization of the decay constant $f$, which is related to the total number of the Nambu-Goldstone bosons and hence the broken group $G$ in the ultraviolet.  Therefore, attributing the nonlinear interaction to the broken group $G$ is  misinformed, as the nonlinearity of the interactions arises entirely from infrared physics: the correct soft limit of S-matrix elements in the presence of a linearly realized group $H$.

The new approach advocated in Refs.~\cite{Low:2014nga,Low:2014oga} intersects with the modern S-matrix program of reconstructing effective field theories from soft behaviors of scattering amplitudes \cite{Cheung:2014dqa,Cheung:2016drk,Cheung:2015ota}, as well as the early attempt by Susskind and Frye in this regard \cite{Susskind:1970gf}: by starting from a 4-point (pt) vertex that satisfies the Adler's zero condition, one can construct a 6-pt amplitude using only the 4-pt vertex, which in general will not have the correct soft limit; this necessitates introducing a 6-pt vertex, whose value is determined by the Adler's zero condition in the 6-pt amplitude. This procedure is then repeated recursively. The nonlinear shift symmetry introduced in Refs.~\cite{Low:2014nga,Low:2014oga} is  the concrete realization of the recursive procedure at the Lagrangian level.

More generally, there is a rich history of studying soft behaviors of massless  particles in gauge theories and gravity \cite{Low:1954kd,GellMann:1954kc,Low:1958sn,Burnett:1967km,Weinberg:1964ew,Weinberg:1965nx,Gross:1968in,Jackiw:1968zza}. The subject gained renewed interests in recent years \cite{ArkaniHamed:2008gz,Strominger:2013lka,Strominger:2013jfa,He:2014cra,Bern:2014vva} and have since found surprising connections to  deep puzzles in different areas \cite{Strominger:2017zoo}. In particular, it is now realized that soft theorems in both gauge theories and gravity are dictated by Ward identities associated with large gauge transformations that do not vanish at the null infinity \cite{Strominger:2017zoo,Hamada:2018vrw}. For Nambu-Goldstone bosons and nonlinear sigma models (\nlsm), soft limits were studied in Refs.~\cite{Kampf:2013vha,Cachazo:2015ksa,Du:2015esa,Low:2015ogb,Cachazo:2016njl,Du:2016njc,Li:2017fsb,Low:2017mlh,Bogers:2018zeg}. In particular, Ref.~\cite{Cachazo:2016njl} computed the subleading single soft limit of Nambu-Goldstone bosons using the Cachazo-He-Yuan (CHY) formulation of scattering amplitudes \cite{Cachazo:2013hca,Cachazo:2013iea,Cachazo:2014xea} and uncovered a mysterious ``extended theory" residing in the coefficient of the subleading single soft limit. The extended theory is a mixed theory containing biadjoint cubic scalars interacting with the Nambu-Goldstone bosons. Only the tree-level amplitudes of the mixed theory are given, but not the Feynman rules. Very little else is known about this mixed theory. It turned out the IR perspective invoking the nonlinear shift symmetry  is ideal for studying the subleading single soft limit of nonlinear sigma models, and some results were summarized in Ref.~\cite{Low:2017mlh}.

In this work we continue the exploration on the infrared structure of the Nambu-Goldstone boson and the nonlinear sigma model. It is organized as follows. In Section \ref{sect:2} we remind the readers of some basic facts of Nambu-Goldstone bosons that are not commonly emphasized in the literature, in particular the importance of vacuum superselection rule and  the special role of soft Nambu-Goldstones in probing the nearby degenerate vacua. Furthermore, we argue that the Adler's zero can be viewed as a consequence of superselection rule.\footnote{Such a viewpoint turns out to be similar to recent efforts in understanding the infrared structure of QED, as well as a proper formulation of its S-matrix elements \cite{Gabai:2016kuf,Mirbabayi:2016axw,Kapec:2017tkm}.}  Then in Section \ref{sect:shiftall} we present the nonlinear shift symmetry to all orders in $1/f$ and the consequence of shift symmetry on the quantum correlation functions, by deriving the associated conserved current and  Ward identity. We also demonstrate how the old-fashioned current algebra can be reproduced from the infrared data, without invoking the coset $G/H$. Soft theorems of \nlsm\ are derived in Section \ref{secnlsmss}, where some subtleties in previous derivations using current algebra are clarified. In Section \ref{secnlsmext} we provide the Feynman rules for the mysterious extended theory of biadjoint scalars in two different parameterizations of \nlsm, and find dramatic simplifications in the Cayley parameterization. The subleading triple soft limit in \nlsm\ is presented in Section \ref{sect:triple}, with the lengthy details of the derivation given in Appendix \ref{appendix}. In particular, we find that the subleading triple soft limit is also controlled by on-shell amplitudes of the same extended theory as in the subleading single soft limit. We conclude and provide an outlook in Section \ref{sect:conclu}.

\section{Soft Nambu-Goldstone Bosons and Degenerate Vacua}
\label{sect:2}

The presence of Nambu-Goldstone bosons is associated with the degenerate vacua. These vacua are related to one another by symmetry operations which leave the dynamics of the physical system invariant. For any given symmetry the corresponding conserved charge is defined from the Noether current ${\cal J}^\mu$
\be
\label{eq:qdefinition}
Q=\int d^3 \vec{x} \ {\cal J}^0(x) \ , \qquad \frac{d}{dt}Q = 0 \ .
\ee
To focus on the essential features we wish to highlight, we will base the following discussion on  a spontaneously broken $U(1)$ group, although it should be clear that our arguments generalize to the more complicated scenarios. Quantum mechanically the unitary operator that represents the action of the symmetry on the Hilbert space of the system is given by
\be
U(\vep)=e^{i Q\vep} \ , \qquad  [U(\vep), H] = 0 \ ,
\ee
where $H$ is the Hamiltonian of the system. If the charge $Q$ annihilates the vacuum $Q|0\rangle=0$, the ground state is invariant under the symmetry transformation: $U(\vep)|0\rangle = |0\rangle$. In this case, there is a unique vacuum, and excited states $\{|n\rangle\}$ that transform into each other under $U(\vep)$ are degenerate in energy:
\be
H \ U(\vep) |n \rangle = U(\vep) \ H|n\rangle = E_n\ U(\vep) |n\rangle\ .
\ee 
In other words, the excited states break up into degenerate multiplets and furnish linear representations of the symmetry group generated by $Q$. The symmetry is manifest in the spectrum of the theory. This is sometimes referred to as the Wigner-Weyl mode of symmetry realization.

If $Q|0\rangle \neq 0$,   one can define
\be
\label{eq:coherent}
|0\rangle_\vep= U(\vep)|0\rangle \ ,
\ee
which has the same energy as $|0\rangle$ and, therefore, is  another ground state.  In this case the degenerate ground states can be labelled by $\vep$. The presence of degenerate vacua does not necessarily imply spontaneous symmetry breaking and massless Nambu-Goldstone bosons. This is because, generally speaking, quantum mechanics allows for tunneling through classical barriers, and if such tunneling were to happen, it could induce mixing among the degenerate states such that a ground state invariant under $U(\vep)$ could be obtained. For example, let's consider the linear combination 
\be
\label{eq:falsevac}
\widetilde{|0\rangle} =\int d{\alpha}\ U(\alpha)|0\rangle \ .
\ee
Then obviously $\widetilde{|0\rangle}$ is invariant under the action of $U(\vep)$ and could be used to construct a theory where the symmetry generated by $Q$ is manifest.

This argument shows that another important ingredient for spontaneous symmetry breaking is the superselection rule \cite{Weinberg:1996kr}: the matrix elements of any local operators ${\cal O}_i$ between different ground states vanish:
\be
\label{eq:superselection}
 _\vep\langle 0 | {\cal O}_i | 0 \rangle_{\vep'} = 0\ .
\ee
As a result, any symmetry-breaking perturbation will  also be diagonal in this basis and yield a ground state that has a definite $\vep$, instead of a mixed state like in Eq.~(\ref{eq:falsevac}). The superselection rule implies the Hilbert space of the theory is divided into disjoint, orthogonal  sectors, each containing a unique vacuum $|0\rangle_\vep$. No local operators can take states from one superselecting sector to another. When this happens the ground state of the system is not  invariant under $U(\vep)$ and the symmetry is spontaneously broken. This is the Nambu-Goldstone phase of symmetry realization.

The superselection rule also explains why the broken symmetry is not manifested in the spectrum like in the Wigner-Weyl mode: states acted upon by $U(\vep)$ belongs to other  superselecting sectors and simply do not exist in the spectrum.

But when does the superselection rule hold? Consider a system in $d$-dimensional spacetime initially in the vacuum $|0\rangle$. Quantum mechanically the probability to create a vacuum bubble of size $R$ in a different ground state $|0\rangle_\vep$ is suppressed by the factor $e^{-S/\hbar}$, where the action is given by the surface tension $T_p$ integrated over the area of the bubble:
\be
S\sim T_p\ R^{d-2}\ .
\ee
For a system of infinite size in $d>2$, to change the vacuum everywhere from $|0\rangle$ to $|0\rangle_\vep$ we let $R\to\infty$ and the tunneling probability now becomes zero, giving rise to the superselection rule. Another way of saying it, the energy cost for the vacuum-to-vacuum tunneling to occur everywhere is infinite for an infinite system. Therefore, strictly speaking, the phenomenon of spontaneous symmetry breaking only occurs for the idealized infinite system in $d>2$. The case of $d=2$ is an exception because the surface of the bubble is just a point and does not grow with the radius. As a result, any long-range order is destroyed by small perturbations and there is no spontaneous symmetry breaking in $d=2$ \cite{Coleman:1973ci}.  

When the symmetry $Q$ is spontaneously broken, the Goldstone theorem states the existence of a massless boson which couples to the broken current ${\cal J}^\mu$. Using relativistically normalized one-particle states,
\be
\label{eq:normalization}
\langle p|k\rangle= (2\pi)^3 (2\omega_{{p}})\, \delta^{(3)}(\vec{p}-\vec{k})\ ,
\ee
Lorentz invariance dictates that
\be
\label{eq:gbdefine}
\langle p|{\cal J}^\mu(x) |0\rangle = i {f \, p^\mu \, e^{ip\cdot x}} \ ,
\ee
where $f$ is the Goldstone decay constant. Note that the current conservation implies $p^2=0$, i.e. the existence of a massless particle. It follows from Eqs.~(\ref{eq:qdefinition}) and (\ref{eq:gbdefine}) that
\be
\langle p| Q | 0\rangle = i f\, \omega_p \, e^{i\omega_p t} \ \delta^{(3)}(\vec{p}) \ .
\ee
Comparison with Eq.~(\ref{eq:normalization}) then reveals that 
\be
Q |0\rangle \sim |\vec{k}=0\rangle + \cdots \ ,
\ee
that is the broken charge $Q$ creates a zero-momentum Nambu-Goldstone boson out of $|0\rangle$, among other things. This is also clear from the usual expression for the axial current in low-energy QCD:
\be
{\cal J}_\mu = f_\pi \partial_\mu \pi + \cdots \ .
\ee
Using free field expansion of the pion,
\be
\pi(x) = \int \frac{d^3 \vec{k}}{(2\pi)^3 2\omega_{\vec{k}}} \left( a_{\vec{k}} e^{-ik\cdot x} + a^\dagger_{\vec{k}} e^{ik\cdot x}\right) \ ,
\ee
the charge operator for the broken symmetry can be expressed as
\be
\label{eq:qzero}
Q=\int d^3 \vec{x}\ {\cal J}_0 = f_\pi \int d^3 \vec{x} \ \partial_t \pi(x) +\cdots = -\frac{i f_\pi}2 (a_{\vec{0}} - a^\dagger_{\vec{0}}) +\cdots \ ,
\ee
which is the creation and annihilation operator for a zero-momentum Nambu-Goldstone boson. Notice that terms omitted above are higher orders in $1/f_\pi$ and contains creation and annihilation operators for $2k+1$, $k\ge 1$, soft Goldstones, as we will see in later sections. Eq.~(\ref{eq:qzero}) gives an interpretation of the degenerate vacuum $|0\rangle_\epsilon$ in Eq.~(\ref{eq:coherent}) as the coherent state of zero-momentum Goldstone bosons.

An interesting corollary of combining the superselection rule with the observation that the broken charge $Q$ creates a zero-momentum Nambu-Goldstone boson is that S-matrix elements containing one soft Goldstone boson vanishes. As is evident by now, external states containing $n$ hard particles and a zero-momentum Goldstone excited over one vacuum can be viewed as excitation of the hard particles over a different nearby vacuum. The superselection rule, which states that no local operators can connect two different vacua, then leads to the vanishing of  such scattering amplitudes. In other words, {\em the Adler's zero is a consequence of the superselection rule of the non-trivial vacua.} Technically speaking,  $Q|0\rangle$ is not a normalizable one-particle state: it is a momentum eigenstate whose position-space wave function is not normalizable in infinite space. This is related to the fact that $Q$ does not exist as a local operator: it generates a {\em global} transformation that flips the vacuum everywhere in the infinite space. The observation reconciles the fact that $Q|0\rangle$ is part of the degenerate vacua with the superselection rule in Eq.~(\ref{eq:superselection}).  On the other hand,  a {\em local} transformation that flips the vacuum in only a finite volume ought to exist -- this is the long wavelength excitation of the Nambu-Goldstone boson: $\pi(x)Q|0\rangle$. It is clear that soft Nambu-Goldstone bosons probe the vacuum structure of the spontaneously broken symmetry.

As we will see later,  the current ${\cal J}^{\mu}$ does more than creating a single Nambu-Goldstone boson out of $|0\rangle$. In fact, it also creates an arbitrary odd number of  Nambu-Goldstone bosons which, to the best of our knowledge, has never been discussed in the literature. These couplings allow us to write down a representation of the on-shell scattering amplitudes of Nambu-Goldstone bosons using the current ${\cal J}^\mu$. Moreover, the extended biadjoint scalar theory discovered in the CHY formalism can be seen to emerge from the matrix elements of the broken current.

\section{The Shift Symmetry, the Conserved Current and the Ward Identities}

\label{sect:shiftall}

In this section we first review the generalized shift symmetry proposed in Refs.~\cite{Low:2014nga,Low:2014oga} and derive the corresponding conserved current, which was presented in Ref.~\cite{Low:2017mlh} without proof. Furthermore, we compute the Ward identities associated with the conserved currents and their commutators, thereby establishing the connection with the classic ``current algebra" approach.

\subsection{The effective Lagrangian from the shift symmetry}
\label{secreview}

The CCWZ formalism \cite{Coleman:1969sm,Callan:1969sn} gives the most general construction of the \nlsm\ based on a coset $G/H$, whose generators satisfy
\bea
\label{eq:liealgebra}
[T^i, T^j] = if^{ijk} T^k, \quad [T^i, X^a] = if^{iab} X^b, \quad [X^a, X^b] = if^{abc} X^c + if^{abi} T^i,\label{eqnlsmcm}
\eea
$T^i$ and $X^a$  the unbroken generators in $H$   and broken generators in $G/H$, respectively. For a symmetric coset we have $f^{abc} = 0$. There exists a Nambu-Goldstone boson $\pi^a$ corresponding to each broken generator $X^a$,\footnote{It is well-known that this is not necessarily the case for spontaneously broken spacetime symmetries \cite{Low:2001bw}. It will be interesting to generalize the shift symmetry approach to spacetime symmetries.}  and $\{\pi^a, a=1,2,\cdots\}$ transform according to a certain linear representation of unbroken group $H$. That is, under an infinitesimal transformation 
\be
\label{eq:lieH}
\pi^a(x) \to \pi^a(x) + i \alpha^i (T^i)_{ab} \pi^b(x) + {\cal O}(\alpha^2) \ ,
\ee
where $\alpha^i$ is  real  and $(T^i)_{ab}$ is the corresponding representation of the group generators. The  CCWZ formalism rests on prior knowledge of the Lie algebra in Eq.~(\ref{eq:liealgebra}), which requires specifying the broken group $G$ in the UV.  Here we briefly review the alternative method proposed in Refs.~\cite{Low:2014nga,Low:2014oga} to derive the effective Lagrangian , which only uses the infrared behavior as input without specifying the coset structure.

To facilitate direct comparison between CCWZ and the shift symmetry approach, we  choose a basis  such that all generators are purely imaginary and anti-symmetric, $(T^i)^T=-T^i$ and $(T^i)^{*}=-T^i$. For real representations, this is automatic, while for  a set of complex scalars $\phi^a(x)$ furnishing a complex representation $R$, we  write $\pi^a(x) = ( {\rm Re}\ \phi^a,  {\rm Im}\ \phi^a)$ and the generators are
\be
\label{eq:antiti}
T^i = \left( \begin{array}{cc}
          \phantom{-}  i\ {\rm Im}\ T^i_R \ & i\ {\rm Re}\ T^i_R \\
          -  i\ {\rm Re}\ T^i_R\ &   i\  {\rm Im}\ T^i_R 
            \end{array}
            \right)  \ .
            \ee
The hermiticity of $T_R$ implies ${\rm Re}\, T_R$ is a symmetric matrix while ${\rm Im}\, T_R$ is an anti-symmetric matrix, from which the anti-symmetricity of $T^i$ follows.   It will be convenient to use the bra-ket notation $|\pi\>$ to denote the Nambu-Goldstone bosons $\pi^a$, and let $T^i | \pi\> \equiv |T^i \pi\>$.

The departing point of the shift symmetry is the soft behavior of on-shell scattering amplitudes of Nambu-Goldstone bosons, which exhibit the Adler's zero condition \cite{Adler:1964um}: the amplitudes  must vanish when one external momentum is taken to zero. This condition is realized if the Lagrangian contains the additional shift symmetry apart from the global symmetry  group $H$. If there is only one flavor of  Nambu-Goldstone boson, the shift is
\bea
\pi \to \pi' = \pi + \vep,\label{eqshiftsr} 
\eea
where $\vep$ is a constant, and the resulting Lagrangian at the two-derivative level is simply
\be
{\cal L} = \frac12 \partial_\mu \pi \partial^\mu \pi \ .
\ee
 For a general unbroken group $H$, there is more than one Nambu-Goldstone boson and the generalization of Eq.~(\ref{eqshiftsr}) now becomes  highly non-linear in $|\pi\rangle$. At the leading order,  the generalized shift symmetry can be written as \cite{Low:2014nga,Low:2014oga}
\bea
| \pi' \> = |\pi \> + F_1 (\mt) | \vep\>\ ,\quad F_1(\mt) = 1 + \sum_{n=1}^{\infty} A_n \mt^n\ ,\label{eqshiftf2}
\eea
where
\bea
\mt \equiv \frac{1}{f^2} |T \pi \> \< \pi T| \ .
\eea
In the above $f$ is analogous to the pion decay constant in the QCD chiral Lagrangian and $A_n$ are numerical constants to be determined later. Eq. (\ref{eqshiftf2}) can be obtained by imposing the condition that when all but $\pi^1$ and $\vep^1$ are set to zero,  the case of a single flavor Goldstone in Eq. (\ref{eqshiftsr})  can be recovered, so that Adler's zero condition is preserved.

The entire CCWZ Lagrangian can be reconstructed in the present approach by focusing on two objects that have well-defined (and simpler!) transformation properties under the shift symmetry, which are the Goldstone covariant derivative and the associated gauge connection
\bea
|\mathcal{D}_\mu \pi \> &\to& | \mathcal{D}_\mu \pi' \> = U |\mathcal{D}_\mu \pi \>\ , \label{eqdmpip} \\
{\cal E}_\mu^i T^i &\to& U ({\cal E}_\mu^i T^i ) U^{-1} +i (\partial_\mu U) U^{-1}  \ ,\label{eq:epip}
\eea
where
\be
\quad U = e^{i u^i(\epsilon, \pi) T^i/f}
\ee
is a nonlinear function of $|\pi\rangle$ and $|\vep\rangle$. Expanding in power series in $1/f$, we write
\begin{align}
 | \mathcal{D}_\mu \pi \> &= F_2 (\mt) |\partial_\mu \pi\> \ , & F_2 (\mt) &= 1 + \sum_{n=1}^{\infty} B_n \mt^n &\label{eqppdm} \\
 u^i ( \pi, \vep) &= \frac{1}{f}\< \pi T^i| F_3( \mt) | \vep\>, & F_3 (\mt) &= \sum_{n=1}^{\infty} C_n \mt^{n-1} &\\
 {\cal E}_\mu^i &= \frac{1}{f^2} \langle \partial_\mu\pi | F_4(\mt)  |T^i\pi\rangle \ , & F_4(\mt)&=\sum_{n=0}^{\infty}{D_n}{\mt^n}&
  \end{align}
By demanding that, under the shift in Eq.~(\ref{eqshiftf2}), $|{\cal D}_\mu\pi\rangle$ and ${\cal E}^i_\mu$ transform according to the prescribed fashion in Eqs.~(\ref{eqdmpip}) and (\ref{eq:epip}) allows one to solve for all but one numerical coefficient, order by order in $1/f$, as long as the following ``Closure condition'' is met:
\bea
\left(T^i \right)_{ab}\left(T^i \right)_{cd}+\left(T^i \right)_{ac}\left(T^i \right)_{db}+\left(T^i \right)_{ad}\left(T^i \right)_{bc} = 0.\label{eqclcd}
\eea
The only undetermined coefficient corresponds to an overall rescaling in the normalization of the decay constant $f$. 

The surprising result is that these numerical coefficients can be solved without specifying the broken group $G$. All that is necessary is the Adler's zero condition and invariance under the unbroken group $H$. As such, the resulting effective Lagrangian is universal for all $G/H'$, where $H'$ contains $H$ as a subgroup, up to an overall normalization of the decay constant $f$. In Refs.~\cite{Low:2014nga,Low:2014oga} closed-form expressions for $F_2$, $F_3$ and $F_4$ are given:
\be
 F_2 (\mt)= \frac{\sin \sqrt{\mt}}{\sqrt{\mt}}\ , \quad F_3 = \frac{i }{\sqrt{\mt}}\tan \left( \frac{\sqrt{\mt}}{2} \right)\ , \quad F_4(\mt)= -\frac{2i}{\mt}\sin^2\frac{\sqrt{\mt}}2 \ ,
\ee
which are sufficient for building up the effective Lagrangian.\footnote{As we obtain the same building blocks $\mathcal{D}_\mu$ and ${\cal E}_\mu$ as in CCWZ, we can write down the interaction between Goldstones and other fields that transform under $H$ as well, e.g. by promoting $\partial_\mu$ to $\partial_\mu +i {\cal E}_\mu$.} The equivalence to the CCWZ formalism can be established with the identification
\bea
\label{eq:idfiab}
\left(T^i \right)_{ab} = -if^{iab}\label{eqopid} \ ,
\eea
in which case  the Closure condition in Eq. (\ref{eqclcd}) is equivalent to the Jacobi identity, thereby allowing $f^{iab}$ to be interpreted as  structure constants  (living in the subspace spanned by $G/H$).\footnote{The identification in Eq.~(\ref{eq:idfiab}) is possible only because we choose a basis such that $(T^i)=-(T^i)^T$ in Eq.~(\ref{eq:antiti}).} For a symmetric coset where $f^{abc}$ vanishes, the knowledge of $f^{ijk}$ and $f^{iab}$ is sufficient to reproduce the entire CCWZ Lagrangian.

In the end, the universal Lagrangian for \nlsm, at the two-derivative order and all orders in $1/f$, is
\be
\mathcal{L} = \frac{1}{2} \< \partial_\mu \pi| \frac{\sin^2 \sqrt{\mt}}{\mt} | \partial^\mu \pi \>.\label{eqnlsmlagep}
\ee
The Lagrangian is dictated by the infrared behavior of the Goldstone scattering amplitudes: 1) the Adler's zero condition and 2) the $H$-invariance, without ever specifying what the broken group $G$ is in the UV. The only undetermined parameter is the overall normalization of the decay constant. \footnote{A construction similar in spirit is also presented in Ref. \cite{Ellis:1970nn}, where $H = SU(2)$ is considered. The three cases where $G = SU(2) \times SU(2)$, $E(3)$ and $SO(3,1)$ correspond to $1/f^2 >0$, $=0$ and $<0$, respectively.}

To dispel any remaining doubts on the universality of Eq.~(\ref{eqnlsmlagep}), let's consider two explicit examples: $SU(2)/U(1)$ and $SU(5)/SO(5)$. The former is the minimal coset containing  a complex Nambu-Goldstone boson $\phi$ charged under the unbroken $U(1)$. For the latter, one can obviously identify several $SU(2)$ subgroups in $SU(5)$ and several $U(1)$ subgroups in $SO(5)$, resulting in many complex Nambu-Goldstone bosons. Denote one of them to be $\Phi$. The universality of Goldstone interactions imply interactions of $\phi$ and $\Phi$ must be identical with each other, which are dictated only by the unbroken $U(1)$ and the Adler's zero condition, up to the normalization of the decay constant $f$. Using the CCWZ formalism to write down the two-derivative interactions for $\phi$ and $\Phi$ we obtain \cite{Low:2014nga},
\bea
\label{eq:su2u1}
{SU(2)/U(1)} &\to& |\partial_\mu \phi|^2-\frac1{3f^2} \left| \phi^*\partial_\mu\phi -\phi\partial_\mu \phi^*\right|^2 + \frac{8}{45f^4}\left| \phi^*\partial_\mu\phi -\phi\partial_\mu \phi^*\right|^2 |\phi|^2\nonumber \\
&& \qquad - \frac{16}{315f^6} \left| \phi^*\partial_\mu\phi -\phi\partial_\mu \phi^*\right|^2 |\phi|^4 + \cdots \ , \\
\label{eq:su5so5}
{SU(5)/SO(5)} &\to& |\partial_\mu \Phi|^2-\frac1{48f^2} \left| \Phi^*\partial_\mu \Phi -\Phi \partial_\mu \Phi^*\right|^2 
+ \frac1{1440f^4}\left| \Phi^*\partial_\mu \Phi -\Phi\partial_\mu \Phi^*\right|^2 |\Phi|^2 \nonumber \\
&& - \frac{1}{80640f^6}\left| \Phi^*\partial_\mu \Phi -\Phi\partial_\mu \Phi^*\right|^2 |\Phi|^4  + \cdots \ .
\eea
The interactions of $\Phi$ become identical to those of $\phi$ after the rescaling of $f\to 4f$ in Eq.~(\ref{eq:su2u1}), as expected from the universality.

\subsection{The shift symmetry to all orders in $1/f$}
\label{secshiftnlsm}

Although the closed-form expressions for $F_i, i=2,3,4$ have been derived previously, the general nonlinear shift $F_1$ was presented without derivation  only recently in Ref.~\cite{Low:2017mlh}. The simplest way to derive $F_1$ is to make use of the universality of Eq.~(\ref{eqnlsmlagep}) and perform a ``matching" calculation into the simplest nontrivial unbroken group of $H=SO(2)\approx U(1)$, which we demonstrate below.

For $H=SO(2)$ there is only one unbroken generator, which in the basis of our choice is simply $T_{ab} = i\epsilon_{ab}$. We use the convention  $\epsilon_{12}=-\epsilon_{21}=1$. Then using $(T^2)_{ab}=\delta_{ab}$ one sees
\be
\mt^n = r^{n-1} \mt \ , 
\ee
where $r=\pi^2/f^2 = \pi^a\pi^a/f^2$. Therefore, for an arbitrary polynomial function
\be
f(\mt)\equiv \sum_{n=0}^\infty c_n \mt^n = c_0\, \mathbb{I} +\sum_{n=1}^\infty c_n r^{n-1}  \mt 
 =c_0\, \mathbb{I} + \frac{f(r)-c_0}{r}\ \mt \ ,
\ee
which leads to  
\beal
\pi^{a\,\prime} &= \pi^a + \vep^a + \frac{F_1(r)-1}{r}\  \mt_{ab}\, \vep^b \ , \\
 {\mathcal{D}}_\mu\pi^a &=  \partial_\mu \pi^a + \frac{F_2(r)-1}{r} \ \mt_{ab}\, \partial_\mu\pi^b \ ,\\
 u^i &= \frac1f \frac{F_3(r)}{r}\pi^a\ T^i_{ab}\ \mt_{bc}\ \vep^c \ .
\end{align}
At this point it is convenient to switch to the complex scalar notation, $\phi=(\pi^1+i\pi^2)/\sqrt{2}$, where
\beal
\phi' &= \phi+ \vep + \frac{F_1(r)-1}{r}\  \frac{\phi^*\vep-\phi\vep^*}{f^2} \phi \ ,\label{eq:phprime} \\
 {\mathcal{D}}\phi &= \partial \phi + \frac{F_2(r)-1}{r} \ \frac{\phi^*\partial \phi-\partial \phi^* \phi}{f^2}\phi \label{eq:defcalDi} \ ,\\
 u&=  F_3(r) \frac{\phi^*\vep - \phi\vep^*}{f} \ . 
\end{align}
and $r=2\phi^*\phi/f^2$. We demand that, after $\phi\to \phi'$, ${\mathcal D}\phi$ transforms covariantly,
\be
{\mathcal D}_\mu\phi'=e^{i u/f}\ {\mathcal D}_\mu \phi  
                         = \left[1+ \frac{i}{f^2} F_3(r)(\phi^*\vep-\phi \vep^*) \right] {\mathcal D}_\mu\phi \ ,\label{eq:mathcalDrhs}
\ee
where ${\cal D}_\mu \phi'$ is computed by plugging Eq.~(\ref{eq:phprime}) into Eq.~(\ref{eq:defcalDi}). In order for Eq.~(\ref{eq:mathcalDrhs}) to hold, one derive the following set of differential equations
\bea
(F_1-1)F_2 -2r F_2' &=&0\\
1+F_2^2(1-2F_1+2r F_1')+2r F_2 F_2' &=& 0 \\
\frac1{r}\left(F_1 + F_2 - 2 F_1 F_2 + 2 r F_2 F_1' + 
 2 r F_2'\right) &=& i F_3
\eea
Multiplying the first equation by $F_2$ leads to
\be
\label{eq:2rf2f2p}
2r F_2F_2'=F_2^2(F_1-1)\ ,
\ee
which can be plugged into the second equation to obtain
\be
F_2^2 = \frac1{F_1-2r F_1'}.
\ee
Plugging back into Eq.~(\ref{eq:2rf2f2p}) allows one to solve for 
\be
F_1(r) = - c_1 \sqrt{r}\ \tan(c_1\sqrt{r}+c_2) \ .
\ee
The boundary condition that $F_1(0)=1$ gives $c_2=\pi/2$ while $c_1$ can be absorbed into the normalization of $f$. So  all three functions can be solved for 
\be
{F}_1(r)= \sqrt{r}\cot\sqrt{r} \ ,\quad
{F}_2(r)=\frac{\sin\sqrt{r}}{\sqrt{r}}\ ,\quad
F_3(r)=\frac{i}{\sqrt{r}} \tan\frac{\sqrt{r}}2 \ . 
\ee
The closed-form expression for $F_1$ is the main result of this subsection.

So in the end, we obtain closed-form expressions, valid to all orders in $1/f$, for the nonlinear shift, the Goldstone covariant derivative and the leading two-derivative Lagrangian 
\bea
 \pi^{a\,\prime} &=&\pi^a + [F_1 (\mt)]_{ab}\  \vep^b\ ,\label{eqshift}\\
  \mathcal{D}_\mu \pi^a &=&[ F_2 (\mt)]_{ab}\ \partial_\mu \pi^b, \\
  {\cal L} &=& \frac12 [F_2(\mt)^2]_{ab} \ \partial_\mu\pi^a \partial^\mu\pi^b \label{eqnlsmlagep1}\ .
\eea 
The  Lagrangian in Eq.~(\ref{eqnlsmlagep1}) is invariant under the general nonlinear shift in Eq.~(\ref{eqshift}). The important feature here is that the effective Lagrangian is obtained, to all orders in $1/f$, using only the infrared data: all that is needed is the linear representation of the unbroken group $H$, under which the Nambu-Goldstone bosons transform, together with the Adler's zero condition. It is not necessary to refer to any broken group $G$. This construction enables us to  derive the Ward identity under the shift symmetry, to all orders in $1/f$, without the precise knowledge of coset space $G/H$.

\subsection{The ``Vector" and ``Axial" Ward identities to all orders in $1/f$}

In this subsection we discuss the conserved currents corresponding to the unbroken and the shift symmetries, respectively. Following the terminology from QCD Chiral Lagrangians, we call the currents for the unbroken symmetry the ``vector currents", while those for the nonlinear shift symmetry the ``axial currents." We also derive the corresponding vector and axial Ward identities. While these objects have been discussed extensively in the context of current algebra in low-energy QCD \cite{Treiman:1986ep},  explicit and closed-form expressions of the vector and axial currents to all orders in $1/f$ have never been discussed in the literature, to the best of our knowledge.

Under the linearly realized, unbroken $H$ symmetry, the Nambu-Goldstones transform as
\be
\label{eq:vectorshift}
\pi^a \to \pi^a+ i\alpha_r (T^r)_{ab} \pi^b \ ,
\ee
from which it is straightforward to derive the corresponding vector current and the Ward identity using the path integral approach. In particular, we need to compute the variation of the ${\cal L}$ in Eq.~(\ref{eqnlsmlagep1}) under Eq.~(\ref{eq:vectorshift}), by promoting $\alpha_r\to \alpha_r(x)$.  However, the pieces that are proportional to $\alpha^r(x)$ must vanish identically since the Lagrangian is invariant under $H$-rotation. Thus we only need to focus on  terms that are proportional to $\partial_\mu\alpha^r(x)$, which can only come from the variation of $\partial_\mu\pi^a$, but not the $F_2^2(\mt)$ term, under Eq.~(\ref{eq:vectorshift}):
\be
\label{eq:vectorallf}
J_\mu^r=  \partial_\mu\pi^a [F_2(\mathcal{T})^2]_{ab} (T^r)_{bc} \pi^c \ , 
\ee
\be
\partial^\mu \langle J_\mu^r(x) \pi^{a_1}(x_1)\cdots \pi^{a_n}(x_n)\rangle = \sum_{j=1}^{n} (T^r)_{a_j b} \delta^{(4)}(x-x_j)\langle \pi^{a_1}(x_1)\cdots \pi^b(x)\cdots \pi^n(x_n)\rangle \ .
\ee
For the axial current, proceeding similarly using the general nonlinear shift in Eq.~(\ref{eqshift}), we arrive at
\be
\label{eq:axialcuf}
  {\cal J}_\mu^a = [F_5(\mt)]_{ab}\ \partial_\mu\pi^b \ , \qquad F_5(\mt) =  \frac{\sin\sqrt{\mt} \ \cos \sqrt{\mt}}{\sqrt{\mt}}\ ,
  \ee
  \be
  \label{eq:axialward}
\partial^\mu \langle {\cal J}_\mu^a(x) \pi^{a_1}(x_1)\cdots \pi^{a_n}(x_n)\rangle = \sum_{j=1}^{n} \delta^{(4)}(x-x_j)\langle \pi^{a_1}(x_1)\cdots [F_1(\mt)]_{a_ja}\cdots \pi^n(x_n)\rangle\ .
\ee

It is instructive to see how the classic current algebra, which embodies the commutators of the group generators in Eq.~(\ref{eq:liealgebra}), can be seen from the current approach. To this end we would like to consider the following operations: variations in the vector and axial currents under the general nonlinear shift in Eq.~(\ref{eqshift}) and under the unbroken group in Eq.~(\ref{eq:vectorshift}), respectively. It is sufficient to work  up to the order of $1/f^2$ to see the desired features. More explicitly, using the bra-ket notation we can write the vector and axial currents
\bea
\label{eq:vectorcurrent}
J_\mu^r &=&  \langle \partial_\mu\pi\, T^r\, \pi\rangle  + \frac{\beta_1}{f^2} \langle \partial_\mu\pi\, T^i\,\pi\rangle \langle \pi \, T^i T^r \,\pi\rangle +{\cal O}(1/f^4) \ ,\\
\label{eq:axialcurrent}
|{\cal J}_\mu\rangle &=& |\partial_\mu \pi\rangle +\frac{\gamma_1}{f^2} |T^i\,\pi\rangle\langle \pi\, T^i\,\partial_\mu\pi\rangle +{\cal O}(1/f^4) \ ,
\eea
where $\beta_1=-1/3$ and $\gamma_1=-2/3$. The shift symmetry at this order is
\bea
\label{eq:shift1storder}
|\pi\rangle&\to & |\pi\rangle + |\vep\rangle +\frac{\rho_1}{f^2} |T^i\,\pi\rangle\langle \pi\, T^i\, \vep\rangle \ , \\
\label{eq:shift1storderderi}
|\partial_\mu \pi\rangle&\to & |\partial_\mu\pi\rangle  +\frac{\rho_1}{f^2} |T^i\,\partial_\mu \pi\rangle\langle \pi\, T^i\, \vep\rangle+\frac{\rho_1}{f^2} |T^i\, \pi\rangle\langle\partial_\mu \pi\, T^i\, \vep\rangle \ , 
\eea
with $\rho_1=-1/3$. 

Before proceeding it will be useful to recall two identities \cite{Low:2014nga} that follow from the Closure condition in Eq.~(\ref{eqclcd}):
\bea
\label{eq:closurecond}
&& |T^i\,\partial_\mu\pi\rangle\langle \pi\, T^i\,\vep\rangle +   |T^i\,\pi\rangle\langle \vep\, T^i\,\partial_\mu \pi \rangle+ |T^i\,\vep\rangle\langle\partial_\mu \pi\, T^i\,\pi\rangle=0 \ , \\
&&\langle \partial_\mu\pi\, T^r T^i\, \pi\rangle \langle \pi \,T^i|+\langle \pi\, T^r T^i \,\pi\rangle \langle \partial_\mu \pi\, T^i|+\langle\pi \,T^r T^i\, \partial_\mu \pi\rangle \langle \pi\, T^i|=0 \ , 
\eea
Plugging Eqs.~(\ref{eq:shift1storder}) and (\ref{eq:shift1storderderi}) into the currents in Eqs.~(\ref{eq:vectorcurrent}) and (\ref{eq:axialcurrent}) and using the preceding two identities, the variations in the vector and axial currents are
\bea
J_\mu^r &\to& J_\mu^r + \langle \partial_\mu \pi\, T^r\, \vep \rangle \ , \\
|{\cal J}_\mu\rangle &\to&|{\cal J}_\mu\rangle +\frac1{f^2}\left[\rho_1 |T^i\,\partial_\mu\pi\rangle\langle \pi\, T^i\,\vep\rangle+ (\gamma_1-\rho_1) |T^i\,\pi\rangle\langle\vep\, T^i\, \partial\pi\rangle +
      \gamma_1 |T^i\,\vep\rangle\langle \pi\, T^i\,\partial_\mu\pi\rangle \right]\nonumber\\
       &=& |{\cal J}_\mu\rangle- \frac1{f^2}\ J_\mu^i\  |T^i\,\vep\rangle  \ ,
\eea       
where we have used the explicit values of the numerical constants. In component form, 
\bea
\label{eq:vectorJshift}
J_\mu^r&\to&  J_\mu^r +{\cal J}_\mu^a\, (T^r)_{ab}\, \vep^b\ , \\
\label{eq:axialJshift}
{\cal J}_\mu^a &\to& {\cal J}_\mu^a -  \frac{1}{f^2}(T^i)_{ab}\, \vep^b\ J_\mu^i \ .
\eea
If we further recall Eq.~(\ref{eq:idfiab}), $(T^i)_{ab}=-if^{iab}$, these two results simply encode the commutators, $[T^r, X^a] = if^{rab}X^b$ and $[X^a, X^b] = if^{abr} T^r$, as expected from current algebra.

In a completely similar fashion, we can also work out the variations of currents under the unbroken $H$-rotation in Eq.~(\ref{eq:vectorshift}):
\bea
J_\mu^r &\to& J_\mu^r -i\alpha_t \,\langle \partial_\mu\pi\, [T^t, T^r]\, \pi\rangle \nonumber\\
  && + \frac{\beta_1}{f^2}\left( -i\alpha_t\, \langle\partial_\mu\pi\,[T^t,T^i]\,\pi\rangle\langle\pi\, T^i T^r\,\pi\rangle - i \alpha_t \,\langle\partial_\mu\pi \,T^i\,\pi\rangle\langle\pi \,[T^t, T^i T^r]\, \pi\rangle\right) 
  \nonumber \\
  &=& J_\mu^r - i \alpha_t \,(if^{trs})\, J_\mu^s \\
|{\cal J}_\mu\rangle &\to&  |{\cal J}_\mu\rangle + i \alpha_t\, |T^t\, \partial_\mu\pi\rangle + \frac{\gamma_1}{f^2} i\alpha_t\, |T^iT^t\,\pi\rangle\langle\pi\, T^i\,\partial_\mu\pi\rangle+ \frac{\gamma_1}{f^2}i\alpha_t\, |T^i\,\pi\rangle\langle\pi\, [T^i, T^t]\,\partial_\mu\pi\rangle\nonumber\\
&=&  |{\cal J}_\mu\rangle + i\alpha_t\, T^t\, |{\cal J}_\mu\rangle \ ,
\eea
which again reflect the corresponding commutators, $[T^r, T^s]=if^{rst}T^t$ and $[X^a, T^r]=if^{arb}X^b$, nicely.

\section{The  Soft Theorems of \nlsm}

\label{secnlsmss}

In this section we study in detail the physical consequences of the axial Ward identity in Eq.~(\ref{eq:axialward}), focusing in particular on its relation to the single soft limit of scattering amplitudes.

\subsection{The Ward identity and the single soft theorem}

\label{secnlsmpi}

We need to use the Lehmann-Symanzik-Zimmermann (LSZ) reduction to extract scattering amplitudes from the correlation functions in the Ward identities. In particular, we perform the operation
\bea
\LI \equiv \left(\frac{i}{\sqrt{Z}}\right)^{n} \int d^4 x\, e^{-i q \cdot x}  \prod_{i=1}^n\ \lim_{p_i^2\to 0}\ \int d^4 x_i\, e^{-i p_i \cdot x_i}\, \DAl_i \ ,
\eea
where $\Box_i$ is the d'Alembertian with respect to $x_i^\mu$, on Eq.~(\ref{eq:axialward}). Note that we have taken the on-shell limit for $\pi^{a_i} $ by imposing $p_i^2 \to 0$, $i = 1, \cdots ,n$. The Fourier transformation with respect to $q$ imposes the momentum conservation condition $q = - \sum_{i=1}^n p_i$. The right-hand side (RHS) of Eq.~(\ref{eq:axialward}) has only $(n-1)$ one-particle poles, thus it will vanish in the on-shell limit. The left-hand side (LHS), on the other hand, can be expanded in $1/f^2$ using the series expansion
 \be
 F_5 (x) = \sum_{k=0}^\infty (-4)^k \frac{x^k}{(2k+1)!} \ ,
 \ee
 and the leading term gives
\bea
  \LI\ \left[ \partial_\mu \<0|\partial^\mu \pi^a (x) \prod_{i=1}^n \pi^{a_i} (x_i) |0\>\right]  = -q^2 J^{a_1 \cdots a_n, a} (p_1, \cdots ,p_n) \ ,
\eea
where $J$ is the scattering amplitude with all but one particle on-shell, defined as
\bea
J^{a_1 \cdots a_n, a} (p_1, \cdots ,p_n) = \langle 0| \pi^a (0) | \pi^{a_1}(p_1) \cdots \pi^{a_n}(p_n)\rangle.
\eea
In the literature such an object was called the ``semi-on-shell" amplitude and first considered by Berends and Giele for Yang-Mills theories \cite{Berends:1987me},  who derived a recursion relation for tree level semi-on-shell amplitudes. At that time the semi-on-shell amplitude serves as the building block for efficient construction of on-shell amplitudes of an arbitrary number of legs. Using Feynman vertices given by the Lagrangian, a Berends-Giele recursion relation for $SU(N)$ \nlsm\ has been proposed in Ref. \cite{Kampf:2013vha}. We are going to propose a different recursion relation following from the axial Ward identity.

\begin{figure}[htbp]
\centering
\includegraphics[height=2.5in]{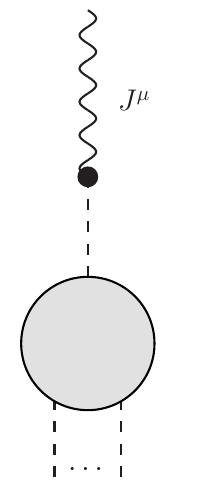}
\qquad
\includegraphics[height=2.5in]{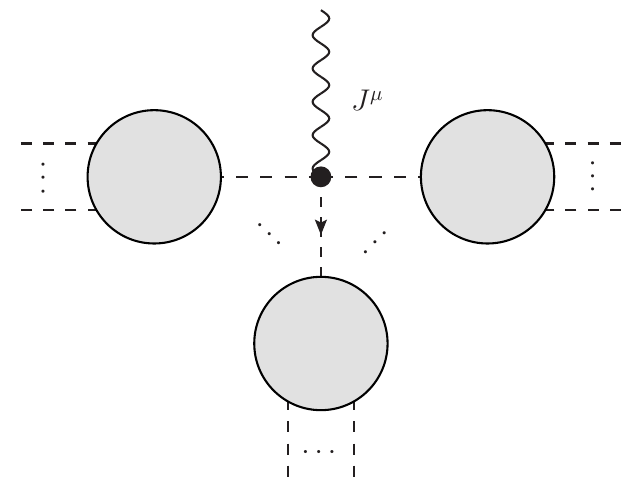}
\caption{\label{figc2kp1} The dashed lines represent Nambu-Goldstone bosons. The current $J_\mu$ can create a one particle pole, as shown on the left. From Eq. (\ref{eqnlsmokd}) we know that the current can also create an odd number of Goldstones, as shown on the right.}
\end{figure}

Terms higher order in $1/f$ on the LHS of Eq.~(\ref{eq:axialward})  have the form 
\be
\label{eq:oakdef}
\<0| \tilde{O}^a_k(q) | \pi^{a_1}(p_1) \cdots \pi^{a_n}(p_n)\>\ ,
\ee
where
\bea
\label{eq:oakexp}
\tilde{O}^a_k(q) &=&  \int d^4 x\, e^{-i q \cdot x} \partial_\mu\left\{ \left[{\cal T}^k(x)\right]_{ab} \partial^\mu \pi^b(x)\right\}  \non\\
& =&i q_\mu \int d^4 x\, e^{-i q \cdot x}  \left[{\cal T}^k(x)\right]_{ab} \partial^\mu \pi^b(x)\ .
\label{eqnlsmokd}
\eea
Observe that $\tilde{O}^a_k$ arises from the $k$-th order term in the axial current in Eq.~(\ref{eq:axialcuf}) and connects the current to $2k+1$ Nambu-Goldstone bosons. While it has been known since the early days of QCD that the axial current creates a one-particle pole, c.f. Eq.~ (\ref{eq:gbdefine}), it is seldom discussed that  the current could also create an odd number of Nambu-Goldstones when higher order effects in $1/f$ are included, as shown in Fig. \ref{figc2kp1}. We will see that this observation plays an important role in trying to connect our results with the mysterious extended theory of biadjoint scalars discovered using the CHY representation of tree amplitudes.

The Ward identity now becomes
\bea
q^2 J^{a_1 \cdots a_n, a} (p_1, \cdots ,p_n)  = \sum_{k=1}^{\infty} \frac{(-4)^k}{(2k+1)!} \<0| \tilde{O}^a_k(q) | \pi^{a_1}(p_1) \cdots \pi^{a_n}(p_n)\> \ .\label{eqnlsmqwi}
\eea
The RHS of Eq. (\ref{eqnlsmqwi}) is proportional to $q^\mu$ because of the form of $\tilde{O}^a_k(q)$ in Eq. (\ref{eqnlsmokd}), so that Adler's zero in $q^\mu\to 0$ is manifest in the present approach, in contrast with the approach using the Feynman vertices in Ref. \cite{Kampf:2013vha}. The $(n+1)$-point on-shell amplitude can be obtained from
\bea
M^{a_1\cdots a_{n+1}} (p_1, \cdots, p_n, p_{n+1})= \lim_{q^2\to 0} -\frac{1}{\sqrt{Z}} q^2 J^{a_1\cdots a_n, a_{n+1}} (p_1, \cdots, p_n)\ ,
\eea
where $p_{n+1} \equiv q = - \sum_{i=1}^n p_i$. Then using Eq. (\ref{eqnlsmokd}) it is easy to see that when $p_{n+1} \to \tau p_{n+1}$, $\tau \to 0$,
\bea
M^{a_1 \cdots a_{n+1}} &=& \tau \frac{1}{\sqrt{Z}} \sum_{k=1}^{\infty} \frac{-(-4)^{k}}{(2k+1)!}  \<0| \int d^4x\, [{\cal T}^k(x)]_{a_{n+1}b}\ ip_{n+1}\cdot \partial\, \pi^b(x) | \pi^{a_1} \cdots \pi^{a_n}\>\non\\
&& + \ordr (\tau^2),\label{eqnlsmssql}
\eea
which is the sub-leading single soft theorem in \nlsm. Notice that at $\ordr(\tau)$, we are able to drop off the exponential $e^{-iq\cdot x}$, so that the operator works like a normal vertex without momentum injection. This is a hint that the matrix elements in the subleading terms  can be interpreted as on-shell scattering amplitudes.

Eq. (\ref{eqnlsmssql}) is valid at quantum level. However, given that we are only including the leading two-derivative Lagrangian, ${\cal O}(p^4)$ terms may be as important as the loop corrections to the two-derivative interactions and should be included for consistency.

\begin{figure}[htbp]
\centering
\includegraphics[width=0.35\textwidth]{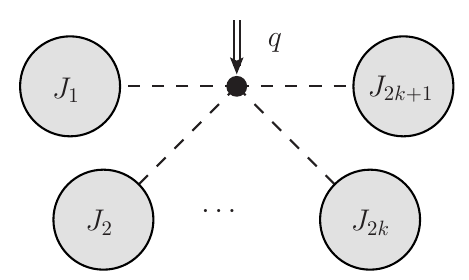}
\caption{\label{figopins} The $(2k+1)$-point vertex with momentum insertion $q$ is connected to $2k+1$ semi-on-shell amplitudes $J_1$, $J_2$, $\cdots$, $J_{2k+1}$.}
\end{figure}

At the classical level, Eq. (\ref{eqnlsmqwi}) becomes a Berends-Giele recursion relation for semi-on-shell amplitudes. The operator $\tilde{{\cal O}}_k^a$ gives rise to a new $(2k+1)$-point vertex
\bea
V^{a_1 \cdots a_{2k+1},a} (p_1, \cdots, p_{2k+1}) = \frac{-i(-4)^{k} }{(2k+1) ! f^{2k}}  \sum_{\sigma \in S_{2k+1}} C^{a_1 \cdots a_{2k+1}, a}_\sigma\ q \cdot p_{\sigma (2k+1)}\ ,\label{eqnlsmfuvt}
\eea
where $\sigma$ is a permutation of $\{ 1,2, \cdots, 2k+1 \}$ and
\bea
 C^{a_1 \cdots a_{2k+1}, a}_\sigma \equiv  T^{i_1}_{a a_{\sigma (1)}} T^{i_1}_{a_{\sigma (2)} b_1} T^{i_2}_{b_1 a_{\sigma (3)}} T^{i_2}_{a_{\sigma (4)} b_2} \cdots  T^{i_k}_{b_{n-1} a_{\sigma (2k-1)}} T^{i_k}_{a_{\sigma (2k)} a_{\sigma (2k+1)}} \label{eqnlsmcf}
\eea
is the color factor associated with the vertex. Notice that  there is a non-zero momentum injection of $q = - \sum_{i=1}^{2k+1} p_i$ at the vertex. Then each term on the RHS of Eq. (\ref{eqnlsmqwi}) can be represented by a Feynman diagram with one insertion of the $(2k+1)$-point vertex connecting either with external legs or internal propagators. But since we are considering only tree amplitudes, each internal propagator must correspond to the off-shell leg of a semi-on-shell (sub)amplitude, as shown in Fig. \ref{figopins}. This is the starting point of constructing a recursive relation for semi-on-shell amplitudes using the axial current, which in turn gives a new representation of on-shell scattering amplitudes where interaction vertices come from expanding the axial current in $1/f$, rather than from the effective Lagrangian in the traditional approach.

Given the above observation, Eq.~(\ref{eqnlsmqwi}) gives the following recursion relation for semi-on-shell amplitudes in \nlsm:
\be
 q^2 J^{a_1 \cdots a_{n}, a} (p_1, \cdots ,p_{n}) = i \sum_{k=1}^{[n/2]} \sum_{l} V^{b_1\cdots b_{2k+1},a}(q_{l^{1}},\cdots,q_{l^{ 2k+1}}) \prod_{i=1}^{2k+1} J^{a_{l^{ i}_1} a_{l^{ i}_2}\cdots, b_i}(\{p_{l^{ i}_j}\})\ ,\label{eqnlsmbgfu}
 \ee
where $l =\{l^{ i}\}$ is a way to divide $\{1,2,\cdots, n\}$ into $2k+1$ disjoint, non-ordered subsets. The $j$-th element of $l^{m}$ is denoted by $l^m_{ j}$ and $q_{l^{m}}=\sum_j p_{l^{m}_{j}}$. Since Eq. (\ref{eqnlsmbgfu}) follows from the axial Ward identity, the Adler's zero in the off-shell leg is manifest. This is not the case for the Berends-Giele recursion relation considered in Ref. \cite{Kampf:2013vha}, which is derived from the Feynman rules of \nlsm. On the other hand, the vertices that connect the semi-on-shell sub-amplitudes in Eq. (\ref{eqnlsmbgfu}) come from the shift symmetry and, as such,  are required to be linear in the momentum $q^\mu$. This  feature gives a very useful handle to extract the subleading single soft factor.

\subsection{Flavor-ordered tree amplitudes and the subleading single soft factor}

Next we consider flavor-ordered amplitudes in \nlsm, which are the main objects studied in Refs. \cite{Kampf:2013vha,Cachazo:2016njl}. Although our formalism so far applies to a general linearly realized symmetry group $H$, flavor-ordered amplitudes can be defined only for certain groups where the product of two traces of group generators can be merged into a single trace. A well-known example is the adjoint of $SU(N)$ group possessing the following property 
\bea
\label{eq:suntrace}
\tr (T^a A) \tr (T^a B) = \tr (AB) - \frac{1}{N} \tr(A) \tr(B)\ ,
\eea
where $A$ and $B$ are two strings of products of group generators. The  disconnected $1/N$ contribution cancels out in the end due to the $U(1)$ decoupling theorem \cite{Elvang:2013cua}. This property is necessary so that the color factors from the subamplitudes in Eq.~(\ref{eqnlsmbgfu}) can be combined into a single trace representing the color factor of the mother amplitude. The discussion in this subsection assumes flavor-ordered amplitudes can be defined properly.

The novelty of building up the \nlsm\ Lagrangian from the IR is so that only generators of the unbroken group $H$ are needed. However, the identification in Eq. (\ref{eqopid}) allows us to ``access" the structure constants related to the shift symmetry, which in the CCWZ formalism correspond to the broken generators $X^a$.  Choosing the normalization $\tr (X^a X^b) = \delta^{ab}$ we can directly translate matrix entries of the unbroken generators into commutators of broken and unbroken generators in CCWZ as follows, 
\be
\label{eqnlsmggtx}
T^i_{ab} = -\tr ([T^i, X^a] X^b),\ , \qquad T^i_{ab} T^i_{cd} = \tr ([[X^a, X^b],X^c]X^d)\ ,
\ee
so that the color factor in Eq. (\ref{eqnlsmcf}) can be written as
\bea
C^{a_1 \cdots a_{2k+1}, a}_\sigma =  {\rm Tr} ( [[\cdots [ [X^a, X^{a_{\sigma (1)}}], X^{a_{\sigma (2)}}], \cdots ], X^{a_{\sigma (2k)}}] X^{a_{\sigma (2k+1)}} )\ .
\eea
It is interesting to observe that the color factor is automatically in the ``half-ladder" basis proposed in Ref.~\cite{DelDuca:1999rs}, which may suggest connections to the color-kinematic duality \cite{Bern:2008qj}.

If we define the flavor-ordered vertex as follows,
\bea
V^{a_1 \cdots a_{2k+1},a} (p_1, \cdots, p_{2k+1}) \equiv \sum_{\sigma \in S_{2k+1}} {\rm Tr} ( X^a X^{a_{\sigma (1)}} \cdots X^{a_{\sigma (2k+1)}} ) V_\sigma(p_1,\cdots,p_{2k+1})\ .
\eea
and denote $V(\mathbb{I}_{n}) \equiv V_{\mathbb{I}_{n}}(p_1,\cdots,p_{n})$, where $\mathbb{I}_{n} \equiv \{ 1,2,\cdots, n \}$ is the identity  permutation, 
then the flavor-ordered vertex arising from the axial Ward identity in Eq. (\ref{eqnlsmfuvt}) is
\bea
V(\mathbb{I}_{2k+1}) = \frac{-i(-4)^k}{(2k+1)!f^{2k}}\sum_{j=0}^{2k} \left(\begin{array}{c}
2k\\
j
\end{array} \right) (-1)^{j} q \cdot p_{j+1}\ ,\label{eqnlsmconv}
\eea
where $q =- \sum_{j=0}^{2k} p_{j+1}$ is the momentum injection at the vertex. Similarly,  the flavor-ordered semi-on-shell amplitudes are defined by
\bea
J^{a_1 \cdots a_{2k+1},a} (p_1, \cdots, p_{2k+1}) \equiv \sum_{\sigma \in S_{2k+1}} {\rm Tr} ( X^a X^{a_{\sigma (1)}} \cdots X^{a_{\sigma (2k+1)}} ) J_\sigma(p_1,\cdots,p_{2k+1})\ .
\eea
Again we use the notation $J(\mathbb{I}_{n}) \equiv J_\sigma(p_1,\cdots,p_{n})$ for $\sigma = \mathbb{I}_{n}$. Furthermore, we choose to normalize one-particle states so that $J(p) = 1$.

In the $SU(N)\times SU(N)/SU(N)_V$ \nlsm\ considered in Ref. \cite{Kampf:2013vha}, the broken generators $X^a$ are isomorphic to the unbroken generators $T^i$ of $SU(N)_V$, in which case Eq. (\ref{eqnlsmggtx})  becomes
\be
T^i_{ab} = -\tr ([T^i, T^a] T^b)\ , \qquad T^i_{ab} T^i_{cd} = \tr ([[T^a, T^b],T^c]T^d).
\ee
Using the identity in Eq.~(\ref{eq:suntrace}) for $SU(N)$ group generators, the recursive relation in Eq. (\ref{eqnlsmbgfu}) for the full amplitudes implies the following Berends-Giele recursion for flavor-ordered amplitudes:
\bea
q^2 J(\mathbb{I}_n)= i \sum_{k=1}^{[n/2]}  \sum_{\{l_m \}} V_{\mathbb{I}_{2k+1}}(q_{l_{1}},\cdots,q_{l_{2k+1}}) \prod_{m=1}^{2k+1} J (l_{m-1}+1, \cdots , l_m)\ ,\label{eqnlsmbgco}
\eea
where $\{l_m \}$ is a  splitting of the ordered set $\{1,2,\cdots, n\}$ into $2k+1$ non-empty ordered subsets $\{l_{m-1}+1,l_{m-1}+2,\cdots,l_m\}$  (here $l_0 = 1$ and $l_{2k+1} = n$). Moreover, $q_{l_{m}} = \sum_{i=l_{m-1}+1}^{l_m} p_i$.

Eq. (\ref{eqnlsmbgco}) involves the flavor-ordered semi-on-shell amplitudes. One can further take the on-shell limit of the off-shell leg of the $(n+1)$-point amplitude on the LHS of Eq. (\ref{eqnlsmbgco}) to obtain a representation of the $(n+1)$-point on-shell  amplitude in terms of the semi-on-shell subamplitudes. In this case the momentum conservation implies
\be
q\cdot p_{2k+1} = q\cdot\left(-q-\sum_{i=0}^{2k-1} p_{i+1}\right) = -\sum_{i=0}^{2k-1} q\cdot p_{i+1} \ ,
\ee
which  allows one to rewrite the vertex in Eq.~(\ref{eqnlsmconv}) as
\be
V (\mathbb{I}_{2k+1}) =  \frac{-i(-4)^k}{(2k+1)!f^{2k}} \sum_{j=1}^{2k-1}\left[ \left(\begin{array}{c}
2k\\
j
\end{array} \right) (-1)^{j} -1 \right] q\cdot p_{j+1}\ .
\ee
One then arrives at  the following expression for the on-shell amplitude,
\bea
M(\mathbb{I}_{n+1}) &=& \sum_{k=1}^{[n/2]}  \frac{-(-4)^k}{(2k+1)!f^{2k}} \sum_{\{l_m \}}\sum_{j=1}^{2k-1} \left[ \left(\begin{array}{c}
2k\\
j
\end{array} \right) (-1)^{j} -1 \right] p_{n+1} \cdot q_{l_{j+1}}\non\\
&&\times \prod_{m=1}^{2k+1} J (l_{m-1}+1, \cdots , l_m)\ ,\label{eqnlsmosex}
\eea
where 
$p_{n+1}=q=-\sum_{i=1}^n p_i$. At this stage Eq. (\ref{eqnlsmosex}) is exact, without taking any soft limit. The single soft limit can be achieved by taking $p_{n+1} \to \tau p_{n+1}$ and $\tau \to 0$, which gives the sub-leading single soft factor of \nlsm\ at tree level. Notice that because the vertex in Eq.~(\ref{eqnlsmosex}) is already linear in $\tau$,  we can drop the $\tau $ dependence in the sub-amplitudes $J (l_{m-1}+1, \cdots , l_m)$ by imposing momentum conservation $\sum_{i=1}^{n} p_i = 0$. The subleading single soft factor was first obtained using the CHY representation of scattering amplitudes in Ref.~\cite{Cachazo:2016njl}, which surprisingly has never been discussed in a field-theoretic context in the vast literature on Nambu-Goldstone bosons and \nlsm. In addition, the coefficient of the order $\tau$ contribution is interpreted in the CHY formalism as the on-shell amplitude of a mysterious extended theory containing cubic biadjoint scalar interactions. We will explore this connection further in the following subsection.

Eqs. (\ref{eqnlsmbgco}) and (\ref{eqnlsmosex}) give very interesting new representations of tree-level semi-on-shell and on-shell scattering amplitudes of Nambu-Goldstone bosons. These representations are different from the traditional Feynman diagrammatic expansion because the vertices involved arise from the axial current in  Eq.~(\ref{eq:axialcuf}), which possesses couplings to an arbitrary odd number of Nambu-Goldstone bosons. As a contrast, recall that for $SU(N)$ \nlsm\ the interaction vertices in the Lagrangian always involve an even number of Nambu-Goldstones.

\begin{figure}[htbp]
\centering
\includegraphics[width=0.25\textwidth]{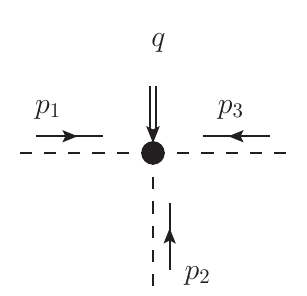}
\qquad
\includegraphics[width=0.25\textwidth]{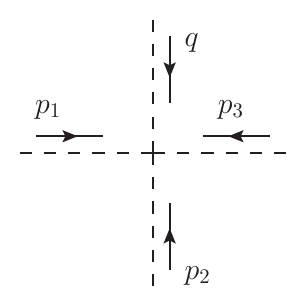}
\caption{\label{fignlsm4p} The two ways to calculate the 3-point semi-on-shell amplitude in \nlsm. On the left we use Eq. (\ref{eqnlsmbgco}) generated by the Ward identity, where we have a 3-point vertex with a momentum insertion $q$. On the right we use the 4-point vertex given by Eq. (\ref{eqnlsmnpve}), which is derived from the Lagrangian.}
\end{figure}

\begin{figure}[htbp]
\centering
\includegraphics[width=\textwidth]{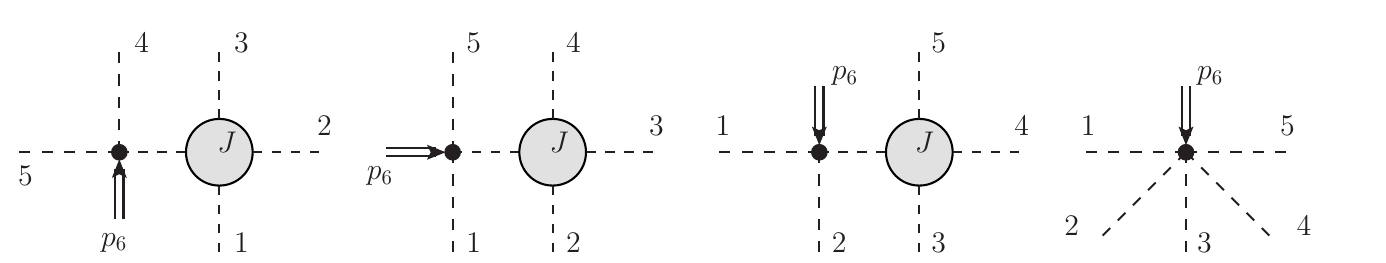}
\\
\includegraphics[width=\textwidth]{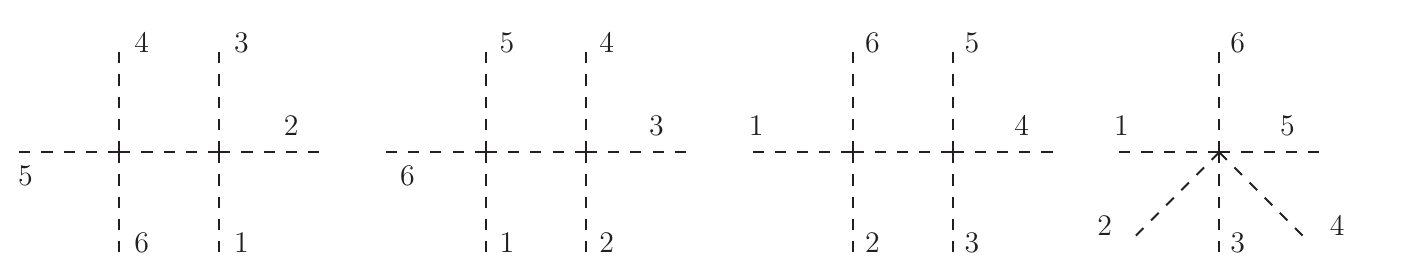}
\caption{\label{fignlsm6p} The diagrams used to calculate the 6-point amplitude of \nlsm. The upper 4 diagrams correspond to using the Ward identity, where we have momentum insertion $p_6$; in three of the diagrams we need to use the result of the 3-point semi-on-shell amplitude. The lower 4 diagrams are the ones generated by Feynman vertices given by Eq. (\ref{eqnlsmnpve}).}
\end{figure}

It is easy to demonstrate the validity of Eqs. (\ref{eqnlsmbgco}) and (\ref{eqnlsmosex}) by comparing with the amplitudes calculated using, for example, the exponential parameterization, which contains the $2n$-point vertices \cite{Kampf:2013vha}:
\be
V^{\text{exp}}_{2n} (p_1, \cdots, p_{2n}) = i \frac{(-1)^{n}}{(2n)!} \left(\frac{4}{f^2}\right)^{n-1} \sum_{k=1}^{2n-1} (-1)^{k-1} \left( \begin{array}{c}
2n-2\\
k-1
\end{array}\right) \sum_{i=1}^{2n} (p_i \cdot p_{i+k}),\label{eqnlsmnpve}
\ee
where $p_{2n+i} \equiv p_i$. For the 4-point amplitude, Eq. (\ref{eqnlsmosex}) gives
\be
M(\mathbb{I}_{4}) =- \frac{1}{f^2} s_{24},\label{eqnlsm4pa}
\ee
where  $s_{ij} \equiv (p_i + p_j)^2$. This is consistent with the amplitude from the 4-point vertex in Eq. (\ref{eqnlsmnpve}). For the 6-point amplitude, we need the 3-point semi-on-shell amplitude, which can be calculated from Eq. (\ref{eqnlsmbgco}):
\bea
J(1,2,3) = - \frac{2}{3 f^2} \frac{(p_1  - 2 p_2   + p_3  ) \cdot (- P_{123} )}{P_{123}^2} =  \frac{1}{f^2} \left( \frac{2}{3}- \frac{s_{12} + s_{23}}{P^2_{123}} \right),\label{eqepj3}
\eea
where $P_{ijk} \equiv p_i + p_j + p_k$. Again, this is the same as the result calculated from using Eq. (\ref{eqnlsmnpve}). The two ways to calculate the 3-point semi-on-shell amplitude are illustrated in Fig. \ref{fignlsm4p}. Then Eq. (\ref{eqnlsmosex}) gives
\bea
M(\mathbb{I}_{6})& =& -\frac{2}{f^2} \left[  p_6 \cdot p_4 J(1,2,3) + p_6 \cdot P_{234} J(2,3,4) + p_6 \cdot p_2 J(3,4,5) \right] \non\\
&& + \frac{2}{3f^4} p_6 \cdot (p_2 - p_3 + p_4) \non\\
&=&- \frac{1}{f^4} \left[ \frac{(s_{12} + s_{23}) (s_{45} + s_{56})}{P^2_{123}} + \frac{(s_{23} + s_{34}) (s_{16} + s_{56})}{P^2_{234}} + \frac{(s_{34} + s_{45}) (s_{16} + s_{12})}{P^2_{345}}\right]\non\\
&&+ \frac{1}{f^4} \left( s_{12} + s_{23} + s_{34} + s_{45} + s_{56} + s_{16} \right),\label{eqnlsm6pa}
\eea
thereby reproducing the well-known $6$-point amplitude in the $SU(N)$ \nlsm. The diagrams corresponding to the new representation of the 6-point amplitude, as well as the traditional Feynman vertices, are shown in Fig. \ref{fignlsm6p}.

At this point it should become clear that the new representation of on-shell amplitudes using the Ward identity amounts to a concrete realization of an old idea proposed by Susskind and Frye in Ref.~\cite{Susskind:1970gf}, where they proposed ``bootstrapping" higher point amplitudes in \nlsm\ from lower point vertices. Amplitudes constructed this way will not satisfy the Adler's zero condition, which necessitates the introduction of higher point vertices. Using the 6-pt amplitude as an illustration, the 5-pt vertex with the current insertion in Fig.~\ref{fignlsm6p} arises from the second-order term in the axial current in Eq.~(\ref{eq:axialcuf}), which owes its existence to fulfilling the Adler's zero condition at $1/f^4$ order.

\subsection{\nlsm\ double-soft theorems}

In this subsection we derive the double soft limit of \nlsm\ from the Ward identities. The double soft limit was first considered in Ref.~\cite{ArkaniHamed:2008gz}, and subsequently derived using current algebra techniques in Ref.~\cite{Kampf:2013vha}. 
Here we wish to not only demonstrate the derivation in the current approach, but also to clarify subtleties that were missed previously.

The double soft limit can be obtained by extending the axial Ward identity in Eq.~(\ref{eq:axialward}) to include the insertion of an additional axial current:
\bea
\label{eq:doubleward}
&& i\partial_x^\mu \langle {\cal J}_\mu^a(x) {\cal J}_\nu^b(y) \pi^{a_1}(x_1)\cdots \pi^{a_n}(x_n)\rangle = \nonumber\\
&&\quad \sum_{j=1}^{n} \delta^{(4)}(x-x_j)\ \langle {\cal J}_\nu^b(y) \pi^{a_1}(x_1)\cdots [F_1(\mt)]_{a_ja}(x)\cdots \pi^{a_n}(x_n)\rangle \nonumber \\
&& \qquad - \delta^{(4)}(x-y) (T^i)_{ab} \ \langle J^i_\nu(x) \pi^{a_1}(x_1)\cdots  \pi^{a_n}(x_n)\rangle \ ,
\eea
where we have used the fact that, under the nonlinear shift symmetry in Eq.~(\ref{eqshift}), the axial current shifts by an amount that is proportional to the vector current, as shown in Eq.~(\ref{eq:axialJshift}). Taking the LSZ reduction on $\pi^{a_i}$ and Fourier-transform with respect to $x$ and $y$, Eq.~(\ref{eq:doubleward}) implies
\bea
\label{eq:doublesofteq}
&&p^\mu q^\nu\ \langle 0| \tilde{\cal J}_\mu^a(p) \tilde{\cal J}_\nu^b(q) | \pi^{a_1}(p_1)\cdots \pi^{a_n}(p_n)\rangle\nonumber\\
&&\qquad= (T^i)_{ab}\ q^\nu\ \langle 0| \tilde{J}_\nu^i(p+q)  | \pi^{a_1}(p_1)\cdots \pi^{a_n}(p_n) \rangle\ .
\eea
Given that each axial current creates a Nambu-Goldstone boson, the LHS of the above equation includes the scattering amplitudes of $n+2$ Goldstone bosons.

\begin{figure}[t]
\centering
\subfloat[]{
\centering
\includegraphics[height=2.5in]{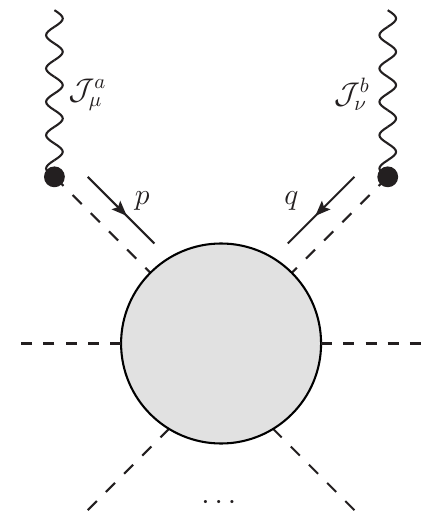}
\label{figdoubleco}
}
\qquad
\subfloat[]{
\centering
\includegraphics[height=2.5in]{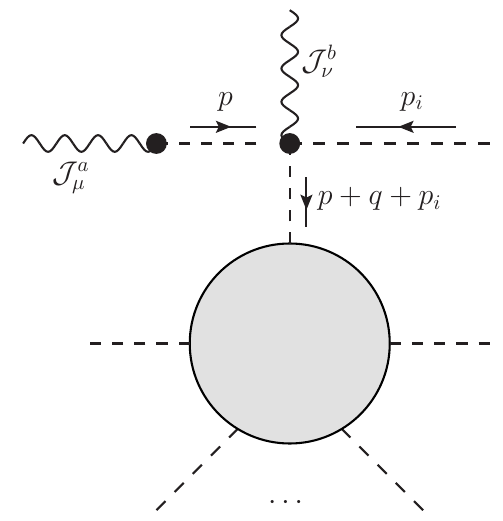}
\label{figdoublecn}
}
\caption{\label{figdoublec} The two contributions to RHS of Eq.~(\ref {eq:doubward}). The first term corresponds to the diagram on the left, while the remainder $R_{\mu\nu}^{ab, \cdots}$ has a non-vanishing contribution in the double-soft limit, which is shown on the right.}
\end{figure}

There is a subtlety, however, that is not taken into account in the previous derivation using current algebra techniques. When factoring out the single Goldstone pole on the LHS of Eq.~(\ref{eq:doublesofteq}), one usually writes
\bea
\label{eq:doubward}
&& \langle 0| \tilde{\cal J}_\mu^a(p) \tilde{\cal J}_\nu^b(q) | \pi^{a_1}(p_1)\cdots \pi^{a_n}(p_n)\rangle\nonumber\\
&&\quad = \sum_{c,d} \frac{i}{p^2} \langle 0|\tilde{\cal J}_\mu^a|\pi^c(p)\rangle\frac{i}{q^2} \langle 0|\tilde{\cal J}_\nu^b|\pi^d(q)\rangle\ \widetilde{J}^{cda_1\cdots a_n}(p,q,p_1,\cdots,p_n) + R_{\mu\nu}^{ab, \cdots}\ ,
 \eea
where $\widetilde{J}^{cda_1\cdots a_n}$ is the $(n+2)$-point amplitude with $c$ and $d$ legs off-shell. The diagram corresponding to the first term on the RHS of Eq.~(\ref {eq:doubward}) is shown in Fig. \ref{figdoubleco}; the remainder $R_{\mu\nu}^{ab, \cdots}$ is  assumed to satisfy the strong regularity conditions \cite{Kampf:2013vha}
 \be
 \lim_{p^\mu \to 0} \lim_{q^\nu \to 0}\ p^\mu q^\nu R_{\mu\nu}^{ab, \cdots} = 0  \ .
 \ee
This turns out to be incorrect.\footnote{To our knowledge $R_{\mu\nu}^{ab, \cdots}$ has never been computed previously in the literature.} Using the all-order expression for the axial current in Eq.~(\ref{eq:axialcuf}), one can compute $R_{\mu\nu}^{ab, \cdots}$ to any order in $1/f$. In particular, as have been emphasized repeatedly, the axial current has non-vanishing matrix elements between $2k+1$ Goldstone bosons and the vacuum, which is given in Eqs.~(\ref{eq:oakdef}) and (\ref{eq:oakexp}). Therefore, there is a class of ``diagrams"  shown in Fig. \ref{figdoublecn} which would contribute to the double soft limit at the leading order,
\bea
\label{eq:rmueq}
 R_{\mu\nu}^{ab, \cdots} &=&  \sum_{i, c,d} \langle 0|\tilde{\cal J}_\mu^a|\pi^c(p)\rangle\ \frac{i}{p^2}\ \langle \pi^c(p)|\tilde{\cal J}_\nu^b|\pi^d(p+q+p_i)\pi^{a_i}(p_i)\rangle\ \frac{i}{(p+q+p_i)^2} \nonumber\\
&& \qquad \times J^{da_1\cdots\hat{a}_i\cdots a_n}(p+q+p_i,p_1,\cdots,p_n) \nonumber \\
&& \qquad+ \left( a \leftrightarrow b, \mu \leftrightarrow \nu\right)  + \widetilde{R}_{\mu\nu}^{ab, \cdots}\ ,
\eea
where $J^{da_1\cdots\hat{a}_i\cdots a_n}(p+q+p_i,p_1,\cdots,p_n)$ is the $n$-point amplitude with the $a_i$-leg omitted and the $d$-leg off-shell. The reason that this is a non-vanishing contribution upon taking the soft limit $p^\mu, q^\nu \to 0$ is that the momentum in propagator in $i/(p+q+p_i)^2$ now goes on-shell. This is the analogy of the ``pole diagram" discussed in Ref.~\cite{Low:2015ogb}. Furthermore, it should be clear from this discussion that the remaining contribution on the RHS of the above equation now satisfies the regularity condition,
\be
 \lim_{p^\mu\to 0} \lim_{q^\nu\to 0}\ p^\mu q^\nu \widetilde{R}_{\mu\nu}^{ab, \cdots} = 0 \ .
 \ee

To compute the double soft limit, we also need to compute the matrix element on the RHS of Eq.~(\ref{eq:doublesofteq}), which can be symmetrized with respect to $(a\leftrightarrow b)$ and $(p\leftrightarrow q)$. Upon using  the expression for the vector current in Eq.~(\ref{eq:vectorallf}), we obtain, in the double soft limit $p^\mu, q^\nu\to 0$,
\bea
&&-\frac{1}2 (T^i)_{ab}\ (p-q)^\nu\ \langle 0| \tilde{J}_\nu^i(p+q)  | \pi^{a_1}(p_1)\cdots \pi^{a_n}(p_n) \rangle\nonumber\\
&& \to- \frac{1}{2f^2} \sum_k (T^i)_{ab}(T^i)_{a_kc} \frac{p_k\cdot(p-q)}{p_k\cdot (p+q)} i M^{a_1\cdots c \cdots a_n}(p_1,\cdots, p_n) \ .
\eea
On the other hand, a similar calculation for $R_{\mu\nu}^{ab, \cdots}$ in Eq.~(\ref{eq:rmueq}) gives
\be
\label{eq:Rcontri}
p^\mu q^\nu R_{\mu\nu}^{ab, \cdots} \to - \frac1{f^2}\sum_k (T^i)_{ab}(T^i)_{a_kc} \frac{p_k\cdot(p-q)}{p_k\cdot (p+q)} i M^{a_1\cdots c \cdots a_n}(p_1,\cdots, p_n) \ .
\ee
In the end, we obtain the double soft limit
\bea
\label{eq:doublesoftfinal}
&& \lim_{p^\mu \to 0} \lim_{q^\nu \to 0} M^{aba_1\cdots a_n}(p,q,p_1,\cdots, p_n) \nonumber\\
&&\qquad = \frac1{2f^2} \sum_k (T^i)_{ab}(T^i)_{a_kc} \frac{p_k\cdot(p-q)}{p_k\cdot (p+q)} M^{a_1\cdots c \cdots a_n}(p_1,\cdots, p_n)\ .
 \eea
 It is worth commenting that, if we hadn't included properly the contribution in Eq.~(\ref{eq:rmueq}), there would have been an extra minus sign in Eq.~(\ref{eq:doublesoftfinal}). It is interesting to note that our result above  actually agrees with Eq.~(E.43) of Ref.~\cite{Kampf:2013vha}, even though the authors didn't include the extra contribution in Eq.~(\ref{eq:rmueq}). This is due to a  missing minus sign on the RHS of their Eq.~(E.38). Correcting the sign would have rendered Eq.~(E.43) with an extra minus sign, which agrees with our calculation without the contribution in  $R_{\mu\nu}^{ab, \cdots}$.

\section{The Mysterious Extended Theory}

\label{secnlsmext}

In Ref.~\cite{Cachazo:2016njl} a mysterious extended theory containing cubic biadjoint scalar interactions is identified as residing in the coefficient of the sub-leading single soft limit of $U(N)$ \nlsm, by using the CHY formalism of the tree-level scattering amplitudes.\footnote{Note that in $U(N)$ \nlsm, the Nambu-Goldstone corresponding to the subgroup $U(1)$ decouples, so that the amplitudes of $U(N)$ \nlsm\ are equivalent to that of $SU(N)$ \nlsm, and we can write down flavor-ordered amplitudes.} Only flavor-ordered on-shell amplitudes are provided in Ref.~\cite{Cachazo:2016njl} and very little else is known regarding the extended theory. More specifically, it was shown that for $p_{n+1} \to \tau p_{n+1}$ and $\tau \to 0$,
\bea
M_{n+1}^{\text{\nlsm}}(\mathbb{I}_{n+1}) = \tau \sum_{i=2}^{n-1} s_{n+1,i}\ M_n^{\text{\nlsm}\oplus \phi^3}(\mathbb{I}_{n}|1,n,i) + \ordr (\tau^2)\ ,\label{eqchynlsm}
\eea
where the label $\text{\nlsm}\oplus \phi^3$ denotes a mixed theory of \nlsm\ and the biadjoint $\phi^3$ theory. The Nambu-Goldstones $\pi^a$ transfrom under the adjoint of $U(N)$, while the biadjoint scalars $\phi^{a\tilde{a}}$ transform under the adjoint of two groups $U(N)$ and $U(\tilde{N})$. In the amplitude $M_n^{\text{\nlsm}\oplus \phi^3}(\mathbb{I}_{n}|1,n,i)$, the ordering $\mathbb{I}_n$ on the left is for $U(N)$, while the ordering on the right is for $U(\tilde{N})$, so that there are $n$ scalars with external momenta $p_1, \cdots , p_n$, where the ones with momenta $p_1$, $p_n$ and $p_i$ are identified as biadjoint scalars. The rest of the legs are Nambu-Goldstones. The CHY formalism gives a prescription to write down the on-shell amplitudes, but neither the Feynman rules nor the Lagrangian of the mixed theory were given. We will study properties of the extended theory in the present approach,   which allows us to  identify the amplitudes of the extended theory, extract information about the interactions between the Nambu-Goldstones and biadjoint scalars, as well as the Feynman rules of $\pi$-$\phi$ interactions. In particular, we explore behaviors of the extended theory under a different parameterization of the \nlsm: the Cayley parameterization, in which we find the Feynman rules of the mixed theory to be dramatically simplified.

\subsection{Feynman rules for the extended theory}
\label{sect:feynext}

We will begin with Eq.~(\ref{eqnlsmosex}),  which exhibits the momentum factors $p_{n+1} \cdot q_{l_{j+1}}$, where $q_{l_{j+1}}$ is the sum of a subset of momenta $\{ p_1, p_2, \cdots ,p_n \}$. This gives rise to the $s_{n+1,i}$ factors in Eq.~(\ref{eqchynlsm}). The coefficient of $p_{n+1} \cdot q_{l_{j+1}}$ can be interpreted as several \nlsm\ subamplitudes connecting to a special vertex with an odd number of legs. This odd-number vertex does not belong to the \nlsm, having factored out the derivatives into the kinematic invariant $s_{n+1,i}$. That is, the vertex contains no derivatives. We can easily extract the analytic form of the vertex from Eq.~(\ref{eqnlsmosex}), which  is non-zero only for $2\le j \le 2k$ and also invariant under $j \leftrightarrow 2k+2-j$,
\bea
V^{\text{\nlsm} \oplus \phi^3} (\mathbb{I}_{2k+1} | 1,2k+1,j)  =\frac{i}{2} \frac{-(-4)^k}{(2k+1)!f^{2k}} \left[ \left(\begin{array}{c}
2k\\
j-1
\end{array} \right) (-1)^{j-1} -1 \right]\ . \label{eqmixedfr}
\eea
These observations led one to {\em postulate} that the three legs in the vertex, the first, the $(2k+1)$-th and the $j$-th,  carry some special properties that are different from the rest of the vertex, which are Nambu-Goldstone bosons. Given that the vertex carries no Lorentz index, it is natural to designate these three special legs as ordinary scalars, as suggested by the $\phi$ label in Eq.~(\ref{eqmixedfr}). In particular, setting $k=1$ so that all legs in the 3-point vertex are special and none is the Goldstone boson, we have
\bea
V_3^{\phi^3} (1,2,3|1,3,2) = iM_3^{\text{\nlsm} \oplus \phi^3} (1,2,3|1,3,2)=-\frac{i}{f^2},\label{eqbiac}\ ,
\eea
where we have also made it clear that the 3-point vertex is the only diagram contributing to the three-point amplitude. The vertex indicates the existence of cubic interactions among $\phi$. Extending to arbitrary $k$ gives rise the first conjecture:
\begin{itemize}

\item There exist flavor-ordered $(2k+1)$-point vertices involving three $\phi$'s, where two of the $\phi$ fields are adjacent to each other, while the rest of the vertices are Nambu-Goldstone bosons. The vertex Feynman rules  are given by Eq.~(\ref{eqmixedfr}).

\end{itemize}

\begin{figure}[t]
\centering
\includegraphics[width = \textwidth]{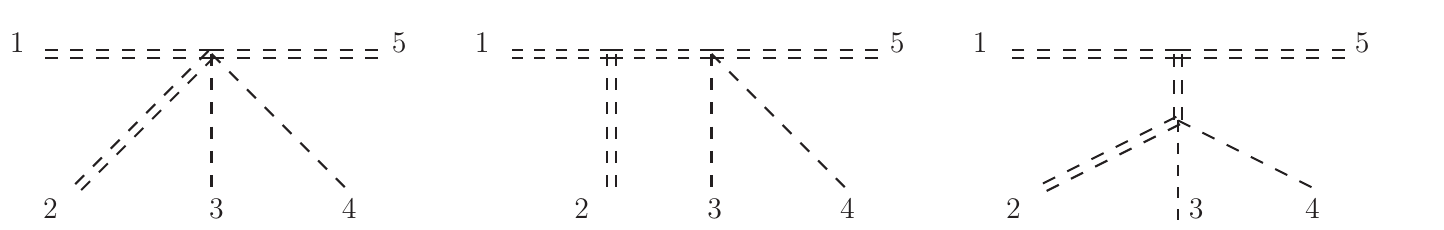}
\caption{The three diagrams representing the contributions to the coefficient of $s_{62}$ in Eq.~(\ref{eq:m6nlsm}), where the single dashed lines represent \nlsm\ Goldstones, while the double dashed lines represent the biadjoint scalars. For the coefficient of $s_{64}$ the diagrams are similar with the appropriate modifications. \label{eqnlsmp3125}}
\end{figure}

\begin{figure}[t]
\centering
\includegraphics[width = 0.7\textwidth]{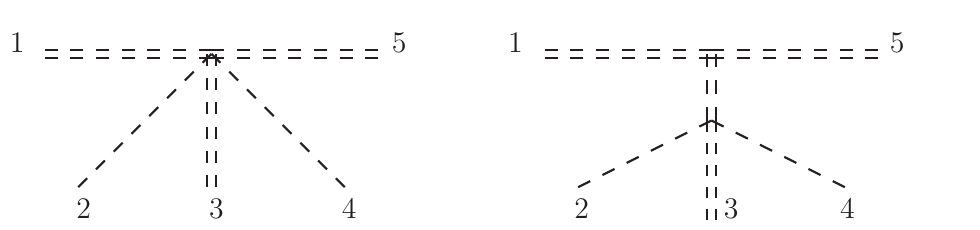}
\caption{The two diagrams  representing the contributions to the coefficient of $s_{63}$ in Eq.~(\ref{eq:m6nlsm}).\label{eqnlsmp3135}}
\end{figure}

Next we study the 5-point amplitude of the extended theory, which would come from taking the single soft limit of the 6-point \nlsm\ amplitude. From Eq.~(\ref{eqnlsmosex}) it is straightforward to compute the subleading single soft limit:
\bea
\label{eq:m6nlsm}
&& iM_{6}^{\text{\nlsm}}(123456)= s_{64}\ V^{\text{\nlsm} \oplus \phi^3}(r45|r54)\ J(123) +s_{62} \ V^{\text{\nlsm} \oplus \phi^3}(12r|1r5) \ J(345) \nonumber \\
&&\quad + (s_{62}+s_{63}+s_{64})\ V^{\text{\nlsm} \oplus \phi^3}(1r5|15r)\ J(234)
         + \sum_{j=2}^4  s_{6j}\ V^{\text{\nlsm} \oplus \phi^3}(12345|15j) \ ,
\eea     
where $r$ represents the flavor index of the internal line connecting to the \nlsm\ subamplitudes $J(ijk)$. The coefficients of the kinematic invariant $s_{6i}, i=2,3,4$ are then
\bea
s_{62}&:& V^{\text{\nlsm} \oplus \phi^3}(12r|1r5) J(345) + V^{\text{\nlsm} \oplus \phi^3}(1r5|15r) J(234) + V^{\text{\nlsm} \oplus \phi^3}(12345|152) , \qquad\\
s_{63}&:& V^{\text{\nlsm} \oplus \phi^3}(1r5|15r) J(234) + V^{\text{\nlsm} \oplus \phi^3}(12345|153) , \\
s_{64}&:&  V^{\text{\nlsm} \oplus \phi^3}(r45|r54) J(123) + V^{\text{\nlsm} \oplus \phi^3}(1r5|15r) J(234) + V^{\text{\nlsm} \oplus \phi^3}(12345|154) . 
\eea
These contributions can be represented diagrammatically in Figs.~\ref{eqnlsmp3125} and  \ref{eqnlsmp3135}. The important observation here is the assignment of the $\phi$-particle flow is essentially dictated by that of the 5-point vertex. Otherwise the various diagrams would not contribute coherently. More specifically, assigning particle 5 to be the $\phi$ particle in the 5-point vertex forces the same choice upon the other diagrams. These observations lead to the second conjecture:

\begin{itemize}

\item There exist flavor-ordered $2k$-point vertices involving two $\phi$'s and $(2k-2)$ Nambu-Goldstones. They have the same Feynman rules as the $2k$-point vertices in the \nlsm.

\end{itemize}

Using these two conjectures, together with the vertex Feynman rule in Eq.~(\ref{eqmixedfr}), we can compute flavor-ordered 5-point amplitudes in the extended theory. For example, the amplitudes corresponding to Figs.~\ref{eqnlsmp3125} and  \ref{eqnlsmp3135} are 
\bea
M_5^{\text{\nlsm} \oplus \phi^3}(\mathbb{I}_{n}|1,5,2) &=& \frac{1}{3f^4} -  \frac{1}{f^4} \left(\frac{2}{3} - \frac{s_{34} + s_{45}}{P^2_{345}} \right) -  \frac{1}{f^4} \left(\frac{2}{3} - \frac{s_{23} + s_{34}}{P^2_{234}} \right),\non\\
&=& \frac{1}{f^4} \left( \frac{s_{23} + s_{34}}{P^2_{234}} +\frac{s_{34} + s_{45}}{P^2_{345}} -1\right)\ , \\
M_5^{\text{\nlsm} \oplus \phi^3}(\mathbb{I}_{n}|1,5,3) &=& \frac{1}{f^4} \left( \frac{s_{23} + s_{34}}{P^2_{234}}  -1\right) \ .
\eea
These results are consistent with Ref. \cite{Cachazo:2016njl}, up to structure constants and minus signs which can be absorbed into a redefinition of the flavor-ordering operation.

Obviously one can continue to build up higher point vertices and amplitudes in the extended theory this way, and it then becomes clear that only odd-number vertices containing three $\phi$'s and even-number vertices containing two $\phi$'s appear in the subleading single soft limit of \nlsm. In particular, no vertices containing a single $\phi$ are ever present. This motivates the third conjecture:

\begin{itemize}

\item No vertices containing only one $\phi$ exist.

\end{itemize}
The importance of this last conjecture is that it implies $\phi$ carries some sort of conserved charge that the Nambu-Goldstones do not, which is consistent with the CHY interpretation of $\phi$ being charged under a different flavor group $U(\tilde{N})$.

 In the end, these three conjectures lead to the following expression for the $n$-point amplitudes in the extended theory
\bea
M_n^{\text{\nlsm}\oplus \phi^3}(\mathbb{I}_{n}|1,n,i) &=& \frac{1}{2}\sum_{k=1}^{[n/2]}  \frac{-(-4)^k}{(2k+1)! f^{2k}} \sum_{j=1}^{2k-1} \sum_{\substack{\{l_m \} \\ l_{j}< i \le l_{j+1}}} \left[ \left(\begin{array}{c}
2k\\
j
\end{array} \right) (-1)^{j} -1 \right]  \non\\
&&\times \prod_{m=1}^{2k+1} J (l_{m-1}+1, \cdots , l_m) \label{eq:mexpression}
\eea
where $n$ is odd and $1<i<n$.
We have checked up to 7-point amplitudes that Eq.~(\ref{eq:mexpression}) agrees with those in Ref. \cite{Cachazo:2016njl}.

\subsection{Cayley parameterization}
\label{sect:cayley}

It is well-known that there are many different ways to parameterize the effective Lagrangian of \nlsm. The derivation using the nonlinear shift symmetry in Section \ref{sect:shiftall} coincides with that of the CCWZ. Although on-shell amplitudes are independent of the particular parameterization of the Lagrangian, the Feynman rules and the semi-on-shell amplitudes, however, do depend on the parameterization. For $SU(N)$ \nlsm\ the two-derivative Lagrangian in a general parameterization can be written as
\bea
\mathcal{L} = \frac{f^2}{8}  \tr \left(\partial_\mu U \partial^\mu U^{-1} \right),\label{eqnlsmlaggp}
\eea
where the matrix $U$ has the general form of
\bea
U = 1 + \frac{2i}{f} \pi^a T^a + \sum_{k=2}^{\infty} a_k \left(\frac{2i}{f} \pi^a T^a \right)^k.\label{eqnlsmgnrp}
\eea
In the above $a_k$ are constants that satisfy the constraint $U^{\dagger} U = 1$, and the different choices of $\{a_k\}$ give different parameterizations \cite{Kampf:2013vha}. The exponential (CCWZ) parameterization corresponds to $U = \exp (2i\pi^a T^a/f)$, and Eq. (\ref{eqnlsmlagep1}) is equivalent to Eq. (\ref{eqnlsmlaggp}) when we recognize $\tr (T^a T^b) = \delta^{ab}$ and $[T^a, T^b] = -(T^{c})_{ab} T^c$. Below we will consider the Cayley parameterization
\bea
U= \frac{1 + i\pi^a T^a/f}{1-i\pi^a T^a/f} = 1 + 2 \sum_{k=1}^{\infty} \left(\frac{i}{f} \pi^a T^a \right)^k,
\eea
which gives the flavor-ordered Feynman vertices for \nlsm:
\bea
V(\mathbb{I}_{2k+1}) = 0, \qquad V(\mathbb{I}_{2k+2}) = \frac{i(-1)^k}{f^{2k}} \left( \sum_{i=0}^k p_{2i+1} \right)^2\ .
\label{eq:cayleyvec}
\eea
Notice that in $V(\mathbb{I}_{2k+2})$ only momenta carried by the odd-number legs enter.

\begin{figure}[t]
\centering
\subfloat[]{
\centering
\includegraphics[height=1.5in]{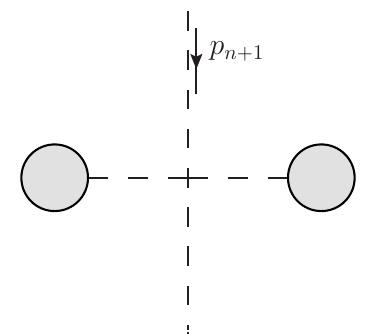}
\label{figcayleya}
}
\qquad
\subfloat[]{
\centering
\includegraphics[height=1.5in]{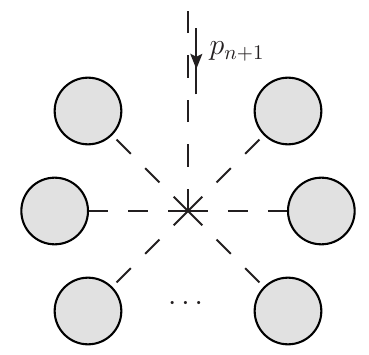}
\label{figcayleyb}
}
\qquad
\subfloat[]{
\centering
\includegraphics[height=1.5in]{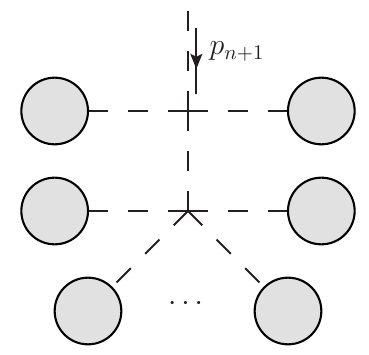}
\label{figcayleyc}
}
\caption{\label{figcayley} A convenient classification of Feynman diagrams of \nlsm\ in the Cayley parameterization. The soft leg with momentum $p_{n+1}$ is labeled, while the blobs represent semi-on-shell sub-amplitudes.}
\end{figure}

When writing down the tree-level flavor-ordered amplitudes, $M_{n+1}^{\text{\nlsm}}$, using Feynman diagrams, the presence of the Adler's zero when $p_{n+1} \to 0$ is usually not explicit, and will be obtained only after summing over all diagrams contributing to the amplitude. In the Cayley parameterization, however, there is a class of diagrams which vanishes in the limit $p_{n+1} \to 0$ on its own. This class consists of diagrams containing a 4-point vertex whose 1st and 3rd legs are the soft leg and an external leg, respectively, as shown in Fig. \ref{figcayleya}. In this case the contribution to the amplitude from this class of diagrams is proportional to the 4-point vertex,
\be
V_4= \frac{-i}{f^2}( p_{n+1}+p_i)^2 \ , \qquad i \neq n+1 \ ,\label{eqcl4p}
\ee
which vanishes as $p_{n+1} \to 0$. This implies the remaining Feynman diagrams contributing to the same amplitude must also vanish in the single soft limit.

These remaining contributions to $M_{n+1}^{\text{\nlsm}}$ can be classified into two classes whose contributions cancel each other in the single soft limit. Recall that in previous sections we have defined $\{l_m \}$ to be a splitting of the ordered set $\{1,2,\cdots, n+1\}$ into $2k+1$ non-empty ordered subsets, with $q_{l_{m}}=\sum_{j=l_{m-1} + 1}^{l_m} p_{j}$. For each $\{l_m\}$, there is a class of diagrams where the soft leg $p_{n+1}$ is connected to a $(2k+2)$-point vertex with $k\ge 2$, as shown in Fig. \ref{figcayleyb}, whose amplitude is
\be
\frac{(-1)^k}{f^{2k}} \left( p_{n+1} +\sum_{j=1}^k q_{l_{2j}} \right)^2  \ \prod_{m=1}^{2k+1} J (l_{m-1}+1, \cdots , l_m)\ .
\ee
The second class consists of diagrams where $p_{n+1}$ enters into a 4-point vertex in such a way that the ``opposite" leg is an internal propagator connecting to a $2k$-point vertex, as shown in Fig. \ref{figcayleyc}. The amplitude from these diagrams is
\be
 \frac{-1}{f^2}  \left( p_{n+1} + \sum_{i=2}^{2k} q_{l_{i}} \right)^2\ \frac{i}{\left( \sum_{i=2}^{2k} q_{l_{i}} \right)^2} \frac{i(-1)^{k-1}}{f^{2k-2}}  \left( \sum_{j=1}^k q_{l_{2j}} \right)^2  \  \prod_{m=1}^{2k+1} J (l_{m-1}+1, \cdots , l_m)\ .
\ee
Then one sees that, in the limit $p_{n+1} \to \tau p_{n+1}$, $\tau \to 0$, the ${\cal O}(\tau)$ terms in these two contributions cancel each other and the Adler's zero is manifest.

Adding the three classes of contributions from Fig. \ref{figcayley}, the $(n+1)$-point flavor-ordered \nlsm\ amplitude can be written as
\bea
M_{n+1}^{\text{\nlsm}} (\mathbb{I}_{n+1})& =& \sum_{j=2}^{n-1} \frac{-1}{f^2} \left(p_{n+1} + p_j \right)^2 J (1, \cdots , j-1) J (j+1, \cdots ,n) \nonumber\\
&&+
\sum_{k=2}^{[n/2]} \frac{(-1)^k}{f^{2k}} \sum_{\{l_m \}} \left( p_{n+1} +\sum_{j=1}^k q_{l_{2j}} \right)^2   \prod_{m=1}^{2k+1} J (l_{m-1}+1, \cdots , l_m)\non\\
&&+ \sum_{k=2}^{[n/2]} \frac{-1}{f^2} \sum_{\{l_m \}} \left( p_{n+1} + \sum_{i=2}^{2k} q_{l_{i}} \right)^2 \frac{i}{\left( \sum_{i=2}^{2k} q_{l_{i}} \right)^2} \frac{i(-1)^{k-1}}{f^{2k-2}}  \left( \sum_{j=1}^k q_{l_{2j}} \right)^2  \non\\
&&\qquad \times \prod_{m=1}^{2k+1} J (l_{m-1}+1, \cdots , l_m)\ , \label{qeqnlsmclfd}
\eea
where the three contributions on the RHS of Eq.~(\ref{qeqnlsmclfd}) correspond to the three classes of diagrams in Fig. \ref{figcayleya}, \ref{figcayleyb} and \ref{figcayleyc}, respectively. Eq.~(\ref{qeqnlsmclfd}) can be directly simplified to
\be
\label{eqnlsmclim}
 M_{n+1}^{\text{\nlsm}} (\mathbb{I}_{n+1})  
 = \sum_{k=1}^{[n/2]} \frac{(-1)^k}{f^{2k}} \sum_{\{l_m \}} 2 p_{n+1} \cdot   \sum_{j=1}^k q_{l_{2j}}  \prod_{m=1}^{2k+1} J (l_{m-1}+1, \cdots , l_m) \ ,
\ee
which is explicitly linear in $p_{n+1}$. Note that just like Eq. (\ref{eqnlsmosex}), this is an exact result. However, the semi-on-shell amplitudes $J$'s in Eq. (\ref{eqnlsmosex}) are not the same as ones in Eq. (\ref{eqnlsmclim}), as $J$'s depend on the parameterization.
When we take the single soft limit $p_{n+1} \to \tau p_{n+1}$, $\tau \to 0$, Eq. (\ref{eqnlsmclim}) automatically gives us the subleading single soft limit in the Cayley representation of \nlsm. By using the same arguments as in Sect.~\ref{sect:feynext}, we again arrive at the same three conjectures, except that the vertex Feynman rule with three $\phi$ fields now is dramatically simpler:
\bea
V^{\text{\nlsm} \oplus \phi^3} (\mathbb{I}_{2k+1} | 1,2k+1,j)  = \left\{ \begin{array}{ll}
\frac{i(-1)^k}{f^{2k}} & \text{for even } j,\\
0& \text{for odd }j.
\end{array} \right.\label{eqnlsmclfr}
\eea
Again, setting $k=1$, Eq.~(\ref{eqnlsmclfr}) gives the biadjoint cubic interaction as in Eq.~(\ref{eqbiac}). Although the Feynman rule Eq.~(\ref{eqnlsmclfr}) is different from that in Eq.~(\ref{eqmixedfr}), it is straightforward to check that it gives exactly the same $3$-pt to $7$-pt on-shell amplitudes in the mixed theory as Eq.~(\ref{eqmixedfr}).

\section{The Subleading Triple Soft Limit}
\label{sect:triple}

The dramatic simplification of the Feynman rules in the mixed theory and the simple derivation of the subleading single soft limit of the \nlsm\ in the Cayley parameterization allow us to study the subleading triple soft limit of the \nlsm. In Ref.~\cite{Du:2016njc} it is shown using Feynman rules in the Cayley representation  that tree amplitudes in \nlsm\ vanish when an odd number of adjacent legs are taken soft. In this section we compute the subleading triple soft limit of $n$-pt amplitudes in \nlsm\ and show the coefficient is given by the $(n-3)$-pt amplitudes of the same mixed theory as in the subleading single soft case. In order to compute the subleading triple soft limit, we need to first compute the single and double soft limits of semi-on-shell amplitudes, which are both ${\cal O}(1)$ in the soft limit as we will see. Therefore we only need to focus on those contributions which do not vanish in the soft limit for the semi-on-shell amplitudes.

Let's consider the single soft limit first by taking $p_{i}\to \tau p_{i}$ to be soft, $\tau\to 0$, and $p_{n}^2 \neq 0$ to be off-shell. We can draw the same three classes of Feynman diagrams as in Fig.~\ref{figcayley}, while we single out $p_{n+1}$ in Fig.~\ref{figcayley}, here we single out $p_i$. It is easy to convince oneself that, as $\tau\to 0$,  the same cancellation of ${\cal O}(1)$ terms between Fig.~\ref{figcayleyb} and Fig.~\ref{figcayleyc} as in Eq.~(\ref{qeqnlsmclfd}) is still present even when the momentum $p_n$ is off-shell. So the sum of the two diagrams is ${\cal O}(\tau)$ and can be ignored. Fig.~\ref{figcayleya}, on the other hand, gives the leading non-vanishing contribution in the limit $\tau\to 0$ when the momentum $p_n$ is directly across the $p_{i}$ in the 4-pt vertex. Moreover,  in Fig.~\ref{figcayleya} the $p_i$ and $p_n$ legs can be across from each other only when $i$ is an even number,\footnote{The number $n$ is an even number by assumption.} since the sub-semi-on-shell-amplitudes can only have an odd number of on-shell legs. In other words, this diagram is present only if $i$ is an even number. If $i$ is an odd number, the sum of all diagrams is ${\cal O}(\tau)$. The 4-pt vertex in Fig.~\ref{figcayleya} is proportional to $(p_n+p_i)^2=p_n^2+{\cal O}(\tau)$, which is the leading contribution. Therefore we have the conclusion that, in the limit $p_i\to \tau p_i$, $\tau\to 0$, and $p_n^2\neq 0$, 
\be
\label{eq:Jsinglesoft}
J(\mathbb{I}_{n-1}) = \left\{ \begin{aligned} & {\cal O}(\tau)\ , &\textrm{$i$ is odd} \\
   & \frac1{f^2} J(\mathbb{I}_{i-1})\, J(i+1,\cdots, n-1) + {\cal O}(\tau)\ ,\quad &\textrm{$i$ is even}
   \end{aligned}\right. \ .
\ee

The derivation of the double soft limit is similar to the single soft limit, with the additional complication that, when the two soft legs are connecting to the same 4-pt vertex, the internal propagator  connecting to the same 4-pt could potentially become on-shell and contribute as ${\cal O}(1/\tau)$. This is the ``pole diagram" considered in Ref.~\cite{Low:2015ogb}. Before considering the pole diagram, let's first consider the case where no intermediate propagator goes on-shell. We will take $p_i\to \tau p_i$, $p_j\to \tau p_j$, $\tau\to 0$, and $p_n^2\neq 0$. Without loss of generality, we assume $i<j$. The same argument leading to Eq.~(\ref{eq:Jsinglesoft}) now tells us that, if both $i$ and $j$ are even,
\bea
\label{eq:Jdoub1}
J(\mathbb{I}_{n-1}) &\to &\frac1{f^2} J(\mathbb{I}_{i-1})\, J(i+1,\cdots, n-1) \nonumber \\
 &\to& \frac1{f^4} J(\mathbb{I}_{i-1})\, J(i+1,\cdots, j-1)J(j+1,\cdots, n-1)\ ,
 \eea
which can be seen as a two-step process by first taking $p_i$ to be soft, leading to the first line in the above, and then $p_j$ to be soft in $J(i+1, \cdots, n-1)$. The pole diagram could arise when both soft legs are adjacent to each other. Thus, for $1<i=j-1<n-2$, the double soft limit is
\be
\label{eq:Jdoub2}
\frac1{f^2}\left[\frac{p_{j-2}\cdot p_{j-1}}{p_{j-2}\cdot(p_{j-1}+p_j)}-\frac{p_{j}\cdot p_{j+1}}{p_{j+1}\cdot(p_{j-1}+p_j)}\right] J(\mathbb{I}_{j-2}, j+1, \cdots, n-1)\ ,
\ee
while for $i=j-1=1$ it is
\begin{equation}
\frac{1}{f^2}\frac{p_1\cdot p_3}{p_3\cdot(p_1+p_2)} J(3, \cdots, n-1)\ .\label{eqjdse1}
\end{equation}
In general, only one of the two contributions above will be generated, with the exceptional case when $i=j-1=n-2$, where both of Eqs.~(\ref{eq:Jdoub1}) and (\ref{eqjdse1}) are generated. Adding both contributions we obtain, for $i=j-1=n-2$,
\begin{equation}
\frac{1}{f^2}\frac{p_{n-3}\cdot p_{n-1}}{p_{n-3}\cdot(p_{n-2}+p_{n-1})} J(\mathbb{I}_{n-3})\ . 
\end{equation}
All the other contributions are subleading and ${\cal O}(\tau^2)$. These results, both the single and double soft limits, confirm those presented in Ref.~\cite{Du:2015esa}.

To proceed with the calculation of triple soft limit, we write the $n$-point \nlsm\ amplitude using Eq. (\ref{eqnlsmclim}):
\be
\label{eq:eqnlsmclim}
 M_{n}^{\text{\nlsm}} (\mathbb{I}_{n})  
 = \sum_{k=1}^{n/2-1} \frac{(-1)^k}{f^{2k}} \sum_{\{l_m \}} 2 p_{n} \cdot   \sum_{j=1}^k q_{l_{2j}}  \prod_{m=1}^{2k+1} J (l_{m-1}+1, \cdots , l_m) \ .
\ee
We will be taking the following triple soft limit: $p_i\to\tau p_i$ with $i=i_1, i_2, n$, $i_1<i_2<n$ and $\tau\to 0$. Notice that both $p_{i_1}$ and $p_{i_2}$ connect to some sub-semi-on-shell amplitudes and, given our previous result that both the single and double soft limits of semi-on-shell amplitude are ${\cal O}(1)$, it is clear that the triple soft limit of Eq.~(\ref{eq:eqnlsmclim}) starts with ${\cal O}(\tau)$, which also confirms the result in Ref.~\cite{Du:2016njc}. At the subleading order, there are several contributions to be considered. Broadly speaking, we need to compute three cases when 1) all three soft legs are not adjacent to one another; 2) only two soft legs are adjacent to each other; 3) all three soft legs are adjacent to one another. 

For case 1), we have $i_1, i_2\neq 1, n-1$ and $i_2\neq i_1+1$. There are two further subcases here: 1-i) both $i_1$ and $i_2$ are even legs; 1-ii) both $i_1$ and $i_2$ are odd legs. The other possibilities, when $i_1$ is even and $i_2$ is odd or vice versa, are equivalent to the subcase 1-ii) after the cyclic and reverse-ordering invariance of flavor-ordered amplitudes, due to the fact that $n$ is even by assumption. Although far from being obvious, we show in  Appendix \ref{appendix} that contributions in these two subcases start with ${\cal O}(\tau^2)$ due to cancellations of the ${\cal O}(\tau)$ terms within each subcase.

For case 2), we are free to choose $i_1=1$, $i_2$ is even and $i_2>2$, again using the cyclic and reverse-ordering invariance.  From the single and double soft limits of the semi-on-shell amplitudes we see the case when $i_1=1$ is connected to a subamplitude, instead of directly into the same vertex which $p_n$ is connected to, is ${\cal O}(\tau^2)$ and can be neglected. It turned out, as shown in  Appendix \ref{appendix},   cancellations at ${\cal O}(\tau)$ also occur, similar to case 1). However, there is one subtlety here: in the semi-on-shell amplitude $J(\mathbb{I}_{n-1})$ considered in Eq. (\ref{eq:Jsinglesoft}), there is always a propagator $i/p_n^2$ coming from the off-shell leg, which we assume to be $\ordr (1)$. In the triple soft case, there are possibilities where this propagator develops an $\ordr (1/\tau)$ pole, so that the $\ordr (\tau)$ terms in Eq. (\ref{eq:Jsinglesoft}) can become $\ordr (1)$. These pole diagrams give non-vanishing contributions at $\ordr (\tau)$. In this case, the leading contribution to Eq.~(\ref{eq:eqnlsmclim}) in the triple soft limit, for $p_i\to \tau p_i$, $i=1,2k,n$ with $1<k<n/2$, is
\bea
&&M_{n}^{\text{\nlsm}} (\mathbb{I}_{n})  = \frac{\tau }{f^{2}} \left[ \frac{ p_{n} \cdot    p_2}{p_2 \cdot (p_1 + p_n)}- \frac{ p_{n} \cdot    p_{n-1}}{p_{n-1} \cdot (p_1 + p_n)} \right] \sum_{\substack{j=2\\ j \ne 2k, 2k \pm 1}}^{n-1} s_{2k,j}\non\\
 &&\qquad \times\  M_{n-3}^{\text{\nlsm}\oplus \phi^3}(2,3,\cdots,2k-1,2k+1, \cdots,n-1|2k+1,2k-1,j) + \ordr (\tau^2)\ .\qquad\label{eqtsc2}
\eea
Similarly, in case 3), when all three soft legs are adjacent to one another, both $i_1=1$ and $i_2=n-1$ must directly attach to the same vertex as $p_n$ in order to contribute at ${\cal O}(\tau)$, except for the pole diagrams similar to case 2). Then the leading contribution in this case, $p_i\to \tau p_i$, $i=1,n-1,n$, is
\bea
\label{eq:tripeqnlsmclim}
M_{n}^{\text{\nlsm}} (\mathbb{I}_{n}) &=&  \frac{\tau }{f^{2}}\sum_{j=3}^{n-3}  \left\{ \left[ \frac{ p_{n} \cdot    p_2}{p_2 \cdot (p_1 + p_n)}- \frac{ 2p_{n} \cdot   (p_1 + p_{n-1}) }{ (p_1+ p_{n-1}  + p_n)^2} \right] s_{n-1,j} \right.\non\\
&&+ \left[ \frac{ p_{n} \cdot    p_{n-2}}{p_{n-2} \cdot (p_{n-1} + p_n)} -  \frac{ 2p_{n} \cdot   (p_1 + p_{n-1}) }{ (p_1+ p_{n-1}  + p_n)^2} \right] s_{1,j}  \left. + \frac{2 ( p_1 \cdot p_{n-1})  }{ (p_1+ p_{n-1}  + p_n)^2} s_{n,j} \right\}\non\\
 &&\times M_{n-3}^{\text{\nlsm}\oplus \phi^3}(2,3,\cdots,n-2|2,n-2,j) + \ordr(\tau^2).\label{eqtsc3}
\eea
We have checked our subleading triple soft theorems using 6-pt and 8-pt \nlsm\ amplitudes. The detailed derivation is somewhat lengthy and presented in Appendix \ref{appendix}. Here, as an example, we show all the cases of the triple soft limit of the 6-pt amplitude.

Using Eq. (\ref{eq:eqnlsmclim}), we can write down the 6-point amplitude as
\bea
M^{\text{\nlsm}}_6 (\mathbb{I}_6) &=& - \frac{1}{f^2} \left[s_{26} J(3,4,5)+ (s_{26}+ s_{36}+ s_{46} )J(2,3,4) + s_{46} J(1,2,3) \right]\non\\
&& +\frac{1}{f^4}\left( s_{26} + s_{46} \right).\label{eqcl6pj}
\eea
Using the Feynman rule given by Eq. (\ref{eqcl4p}), we have
\bea
J(1,2,3) =  \frac{1}{f^2} \frac{s_{13}}{P^2_{123}},
\eea
which is different from Eq. (\ref{eqepj3}), as we would expect. Then Eq. (\ref{eqcl6pj}) becomes
\bea
M^{\text{\nlsm}}_6 (\mathbb{I}_6) &=& - \frac{1}{f^4} \left[\frac{s_{26} s_{35}}{P^2_{345}}+ \frac{(s_{26}+ s_{36}+ s_{46} ) s_{24}}{P^2_{234}} + \frac{ s_{46} s_{13} }{P^2_{123}}  - s_{26} - s_{46} \right],\label{eq6pcl}
\eea
which is explicitly linear in $p_6$ and consistent with Eq. (\ref{eqnlsm6pa}). For case 1) of the triple soft limit, we have $p_2$, $p_4$ and $p_6$ to be soft, and it is easy to see that $M^{\text{\nlsm}}_6 (\mathbb{I}_6)$ is at $\ordr (\tau^2)$. For case 2) we take $p_1$, $p_4$ and $p_6$ to be soft. We need the 3-point amplitude of the mixed theory, $ M_{3}^{\text{\nlsm}\oplus \phi^3}(2,3,5|5,3,2) = -1/f^2$. Then Eq. (\ref{eqtsc2}) gives
\bea
M^{\text{\nlsm}}_6 (\mathbb{I}_6)  =- \frac{\tau}{f^4} \left[  \frac{s_{26} s_{24} }{ s_{12}+ s_{26} } - \frac{ s_{56} s_{24} }{ s_{15} + s_{56}} \right] + \ordr (\tau^2).\label{eqts6pc2}
\eea
For case 3) we take $p_1$, $p_5$ and $p_6$ to be soft, and from Eq. (\ref{eqtsc3}) we know that
\bea
M^{\text{\nlsm}}_6 (\mathbb{I}_6) &=& -\frac{\tau }{f^{2}}   \left[ \frac{ s_{26} s_{35} }{s_{12} + s_{26} } + \frac{s_{15}s_{36}- (s_{16} + s_{56}) (s_{13} + s_{35}) }{ P^2_{156} }   +  \frac{ s_{46} s_{13} }{s_{45} + s_{46} }  \right]+ \ordr(\tau^2) .\qquad\label{eqts6pc3}
\eea
It is easy to check that Eqs. (\ref{eqts6pc2}) and (\ref{eqts6pc3}) are consistent with Eq. (\ref{eq6pcl}).

\section{Conclusion and Outlook}
\label{sect:conclu}

In this work we presented a systematic study on the consequence of Adler's zero condition on the correlation functions of \nlsm, employing the IR perspective without recourse to a target coset $G/H$. Given that the Nambu-Goldstone boson is the long wave-length excitation over degenerate vacua, it is only natural that their interactions are insensitive to UV physics.\footnote{Such an observation turns out to have important implications for models addressing the naturalness problem of the 125 GeV Higgs boson \cite{Liu:2018vel}. Although there are a variety of composite Higgs models utilizing different coset structures \cite{Panico:2015jxa,Bellazzini:2014yua}, the interaction of the Higgs bosons is actually universal, which provides us a handle for identifying whether the Higgs boson is composite or not.} The only parameter dependent on the broken group $G$ in the UV is the normalization of the decay constant $f$, which essentially counts the number of Nambu-Goldstone bosons from the requirement of a canonically normalized kinetic term in the Lagrangian. It turns out that the presence of non-trivial vacua, together with the constraints from the linearly realized unbroken group $H$, is sufficient to construct the entire effective action of the Nambu-Goldstone bosons and the \nlsm. The degenerate vacua manifest themselves in the Adler's zero condition, which results from a shift symmetry. The shift symmetry can be viewed as exciting the Nambu-Goldstone boson over a nearby degenerate vacuum. Since the dynamics does not depend on the particular vacuum chosen in the Fock space, the Lagrangian and the Hamiltonian is invariant.

An important ingredient in the existence of Nambu-Goldstone bosons is the superselection rule in the space of degenerate vacua: off-diagonal matrix elements of any unitary operator between distinct vacua must vanish. This ingredient, together with the realization that soft Nambu-Goldstone bosons probe nearby degenerate vacuum, suggest an interpretation of the Adler's zero as a consequence of the vacuum superselection rule. Such an interpretation is similar to the viewpoint employed in recent efforts to formulate S-matrix elements in QED by using asymptotic states dressed by a coherent cloud of soft photons \cite{Gabai:2016kuf,Mirbabayi:2016axw,Kapec:2017tkm}. It would certainly be interesting to pursue this connection further.

To study the constraints of Adler's zero condition on quantum correlators of \nlsm, we derived the conserved current and the Ward identity associated with the nonlinear shift symmetry. Upon the LSZ reduction, we obtained a new representation of the on-shell tree-level amplitudes of Nambu-Goldstone bosons. This representation is a concrete realization of the proposal first introduced by Susskind and Frye in Ref.~\cite{Susskind:1970gf}, where they used a bootstrapping procedure to construct higher point amplitudes in \nlsm\ from lower point amplitudes and introduce higher point vertices to guarantee the Adler's zero condition. In particular, building blocks in the representation are matrix elements of increasingly higher order operators in the conserved current corresponding to the shift symmetry. These operators in the current all contain an odd number of Nambu-Goldstone bosons, which imply the current has a non-vanishing matrix element between the vacuum and any odd number of Nambu-Goldstone bosons.  While it is well-known that the ``axial" current can create a pion out of the vacuum, the fact that the axial current can also create $2k+1$ pions is much less discussed, which plays an important role in making sure the S-matrix element possesses the Adler's zero. The Ward identity for the nonlinear shift symmetry naturally leads to the subleading single soft  and double soft  theorems in \nlsm. Along the way we clarify a subtlety related to the non-vanishing matrix element between the current and three Nambu-Goldstone bosons, which was not taken into account in previous derivations of the double soft limit using current algebra techniques.

The ability to compute the subleading single soft limit also allows us to identify the extended theory containing biadjoint cubic scalars interacting with the Nambu-Goldstone bosons. We are able to obtain the Feynman rules of the extended theory and also study its dependence on the different parameterizations of \nlsm. It turned out the Feynman rules for the extended theory simplify dramatically in the Cayley parameterization.

Given the simplification of calculation in the Cayley parameterization, we proceeded to compute the subleading triple soft limit in the \nlsm\ using conventional Feynman diagrams. The triple soft limit of $n$-pt tree amplitudes starts with ${\cal O}(\tau)$ and, at this order, the coefficient of $\tau$ turned out to be related to the $(n-3)$-pt amplitudes of the same extended theory with biadjoint cubic scalars as in the subleading single soft limit. It would obviously be very interesting to see if it is possible to identity the extended theory directly at the Lagrangian level, using techniques in effective field theories such as the soft-collinear effective theory.

There are several more interesting future directions. One has to do with the color-kinematics duality \cite{Bern:2008qj} in the context of \nlsm, which would become the ``flavor-kinematics" duality \cite{Chen:2013fya,Du:2016tbc,Carrasco:2016ldy,Cheung:2016prv}. In the CHY approach, the flavor-kinematics duality is manifest. By comparing our results with those of the CHY's, it became apparent that the second adjoint index carried by the biadjoint scalar is related to the position of the derivative operator in the matrix element of operators in the current, which hints at potential connections between the adjoint index (flavor) and the derivative (kinematics). On a separate front, it is possible to apply the IR approach in this work to effective theories with ``enhanced" Adler's zero, such as the Dirac-Born-Infeld scalar theory and Galileons \cite{Cheung:2014dqa}. According to the CHY proposal, there are also extended theories residing in the subleading single soft limit. It is possible to extract more information on these extended theories in theories with the enhanced Adler's zero, which will be reported elsewhere \cite{Yin:toappear}.

\acknowledgments
This work is supported in part by the U.S. Department of Energy under contracts No. DE- AC02-06CH11357 and No. DE-SC0010143. We thank John R. Ellis for bringing Ref. \cite{Ellis:1970nn} to our attention.

\appendix

\section{Appendix: Derivation of the Triple Soft Limit in \nlsm}
\label{appendix}

\begin{figure}[htbp]
\centering
\includegraphics[height=1.4in]{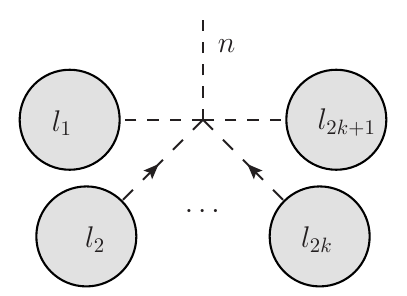}
\caption{\label{figtsste} A diagram from the RHS of Eq. (\ref{eq:eqnlsmclim}). Note that the semi-on-shell sub-amplitudes are labeled by the index of the right most on-shell leg, or the on-shell leg to the left of the off-shell leg. The arrows correspond to momenta that appear in the factor $2p_n \cdot q_{l_{2j}}$. }
\end{figure}

On the RHS of Eq. (\ref{eq:eqnlsmclim}), for a certain $k$ and $\{ l_m\}$ we have a specific diagram shown in Fig. \ref{figtsste}, where we have $2k+1$ semi-on-shell sub-amplitude $J$'s multiplied together. Some of the $J$'s can be a ``one point'' amplitude, which is just an external leg. In this section $J(1) =1$ will be called a ``trivial'' semi-on-shell amplitude; $J$'s with at least 3 on-shell legs are called ``non-trivial''.

We take the triple soft limit of $p_i \to \tau p_i$, $i = i_1, i_2, n$ with $i_1 < i_2 < n$ and $\tau \to 0$. Because the invariance under cyclic permutations of momentum in $M_{n}^{\text{\nlsm}}$, as well as the invariance when we reverse the ordering, we only need to discuss the following 4 cases.

\subsection{Case 1-i)}

In case 1-i), where both $i_1$ and $i_2$ are even, there are 4 scenarios:

\begin{figure}[htbp]
\centering
\includegraphics[width=0.6\textwidth]{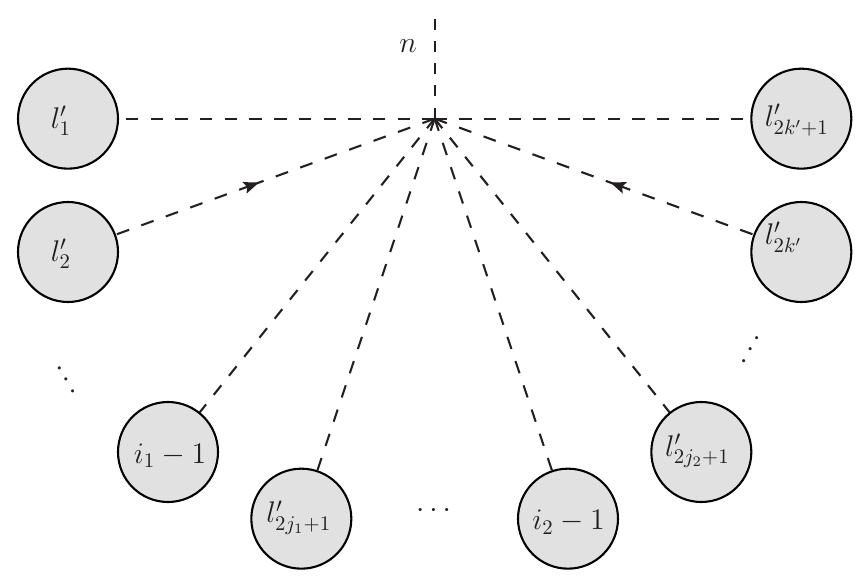}
\caption{\label{figtsc1p1} A diagram when two soft legs $i_1$ and $i_2$ each split a $J$ in half, with even $i_1$ and $i_2$. }
\end{figure}

1. Legs $i_1$ and $i_2$ are attached to different non-trivial $J$'s. By Eq. (\ref{eq:Jsinglesoft}) we know that the contribution is non-zero only if they are both attached to some $J$'s in Fig. \ref{figtsste} labeled by $l_m$ where $m$ is odd. Then they just ``split'' the $J$'s by half, and we arrive at a diagram like Fig. (\ref{figtsc1p1}), where we have $2k+3$ $J$'s multiplied together, so $k'=k$ and our $\{ l_m \}$ is a  splitting of the ordered set $\{1,2,\cdots, i_1-1, i+1,\cdots,i_2-1, i_2+1, \cdots n-1\}$ into $2k+3$ non-empty ordered subsets $\{l_{m-1}+1,l_{m-1}+2,\cdots,l_m\}$  (here $l_0 = 1$ and $l_{2k+3} = n-1$), with $i_1-1,\ i_2-1 \in \{l_m\}$, and the other elements are denoted in order by $l'_m$. Note that the momentum factor will be $2 p_n \cdot \sum_j q_{l'_{2j}}$, where we sum the $q_{l'_m}$'s of even $m$'s.

\begin{figure}[htbp]
\centering
\includegraphics[width=0.6\textwidth]{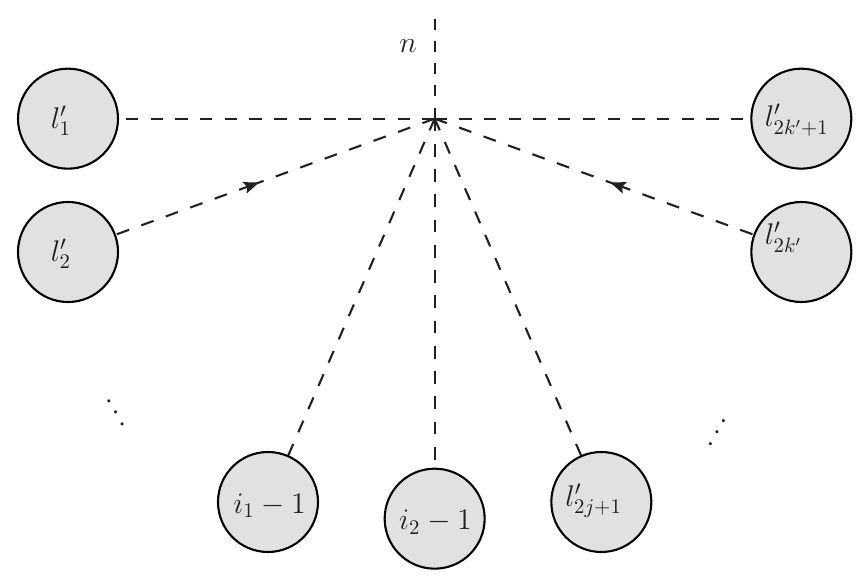}
\caption{\label{figtsc1p2} A diagram when two soft legs $i_1$ and $i_2$ split a $J$ into three parts, with even $i_1$ and $i_2$. }
\end{figure}

2. Legs $i_1$ and $i_2$ are attached to the same non-trivial $J$. From the discussion on double soft limit of semi-on-shell amplitudes in Sec. \ref{sect:triple} we know that the contribution is non-zero only if they are attached to some $J$ in Fig. \ref{figtsste} labeled by $l_m$ where $m$ is odd. Then the two soft legs split the $J$ into three parts, generating the diagram in Fig. \ref{figtsc1p2}. Again, $k'=k$, so that we have $2k+3$ $J$'s, two of which are labeled by $i_1-1$ and $i_2-1$, others labeled by $l'_m$, and the momentum factor is still $2 p_n \cdot \sum_j q_{l'_{2j}}$.

The above two scenarios together actually exhaust all the possible  splittings that satisfies $i_1-1,\ i_2-1 \in \{ l_m\}$, without double-counting. Together they give a contribution
\bea
 \tau \sum_{k=1}^{n/2-3} \frac{(-1)^k}{f^{2k+4}} \sum_{\{l_m \}} 2 p_{n} \cdot   \sum_{j=1}^k q_{l'_{2j}}  \prod_{m=1}^{2k+3} J (l_{m-1}+1, \cdots , l_m) +\ordr (\tau^2). \label{eqtsc1p12}
\eea
Note that the largest value that $k$ can take is $n/2-3$: for $k=n/2-2$ or $k=n/2-1$ we cannot have both soft legs attached to non-trivial $J$'s.

3. One of the soft legs $i_1$ and $i_2$ is attached to a non-trivial $J$, while the other is attached to a trivial $J$, namely, connected to the same vertex as soft leg $n$. For $i_2$ attached to a non-trivial $J$, it split the $J$ by half, generating the same diagram as either Fig. (\ref{figtsc1p1}) or Fig. (\ref{figtsc1p2}). However, this time we only have $2k+1$ $J$'s in the diagram, so that $k' = k-1$ in contrast to the previous two scenarios where we have $2k+3$ $J$'s. For $i_1$ attached to a non-trivial $J$ the situation is similar and the contribution is exactly the same. In the end this scenario gives the contribution
\bea
&&2\tau \sum_{k=2}^{n/2-2} \frac{(-1)^k}{f^{2k+2}} \sum_{\{l_m \}} 2 p_{n} \cdot   \sum_{j=1}^{k-1} q_{l'_{2j}}  \prod_{m=1}^{2k+1} J (l_{m-1}+1, \cdots , l_m) +\ordr (\tau^2)\non\\
&=&-2\tau \sum_{k=1}^{n/2-3} \frac{(-1)^k}{f^{2k+4}} \sum_{\{l_m \}} 2 p_{n} \cdot   \sum_{j=1}^k q_{l'_{2j}}  \prod_{m=1}^{2k+3} J (l_{m-1}+1, \cdots , l_m) +\ordr (\tau^2).\quad\ \label{eqtsc1p3}
\eea
Note that in the first and second line of Eq. \ref{eqtsc1p3} $\{ l_m \}$ split the indices into $2k+1$  and $2k+3$ subsets, respectively; similar treatment is implied in the equations below. We have a factor of $2$ in front because both $i_1$ and $i_2$ can be the one attached to a non-trivial $J$. Also, note that in the first line of Eq. (\ref{eqtsc1p3}), the largest value of $k$ is $n/2-2$, as for $k=n/2-1$ it is impossible to attach any leg to a non-trivial $J$; actually, the only possible diagram there is the $n$ point vertex. Also, the $k=1$ case in the first line is at $\ordr (\tau^2)$. Then Eq. (\ref{eqtsc1p3}) is just $-2$ times Eq. (\ref{eqtsc1p12}).

4. Both of the soft legs $i_1$ and $i_2$ are directly attached to the same vertex as soft leg $n$. Again, we have the same diagrams as either Fig. (\ref{figtsc1p1}) or Fig. (\ref{figtsc1p2}). This time we only have $2k-1$ $J$'s, so that $k' = k-2$, and the contribution is
\bea
&&\tau \sum_{k=3}^{n/2-1} \frac{(-1)^k}{f^{2k}} \sum_{\{l_m \}} 2 p_{n} \cdot   \sum_{j=1}^{k-2} q_{l'_{2j}}  \prod_{m=1}^{2k-1} J (l_{m-1}+1, \cdots , l_m) +\ordr (\tau^2)\non\\
&=&\tau \sum_{k=1}^{n/2-3} \frac{(-1)^k}{f^{2k+4}} \sum_{\{l_m \}} 2 p_{n} \cdot   \sum_{j=1}^k q_{l'_{2j}}  \prod_{m=1}^{2k+3} J (l_{m-1}+1, \cdots , l_m) +\ordr (\tau^2).\quad\ \label{eqtsc1p4}
\eea
Note that in the first line of Eq. (\ref{eqtsc1p4}) the sum of $k$ starts from $k=3$, as diagrams with $k=1$ do not exist, and $k=2$ terms are at $\ordr (\tau^2)$. Then Eq. (\ref{eqtsc1p4}) is the same as Eq. (\ref{eqtsc1p12}), and adding everything together all the $\ordr(\tau)$ contributions cancel out, so that $M_{n}^{\text{\nlsm}} (\mathbb{I}_{n}) = \ordr (\tau^2)$.

\subsection{Case 1-ii)}

\begin{figure}[htbp]
\centering
\includegraphics[width=0.6\textwidth]{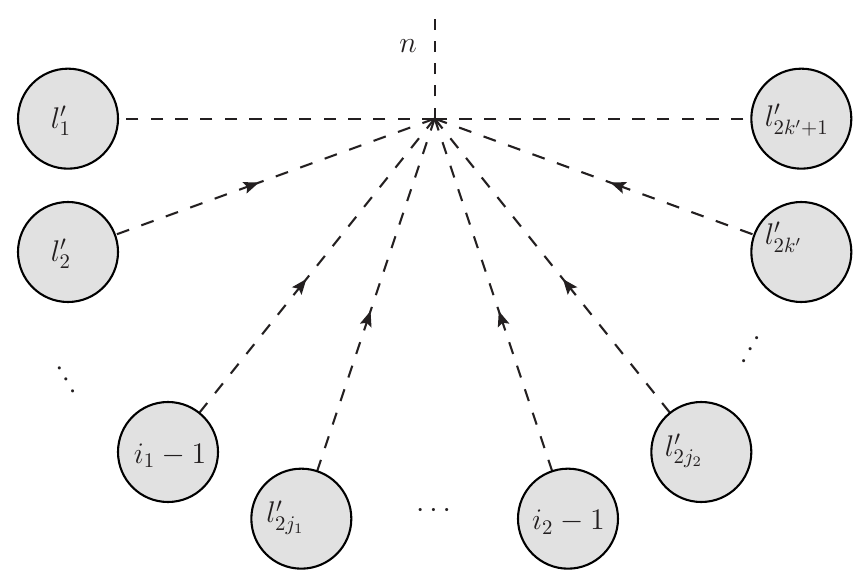}
\caption{\label{figtsc2p1} A diagram when two soft legs $i_1$ and $i_2$ each split a $J$ in half, with odd $i_1$ and $i_2$. }
\end{figure}

\begin{figure}[htbp]
\centering
\includegraphics[width=0.6\textwidth]{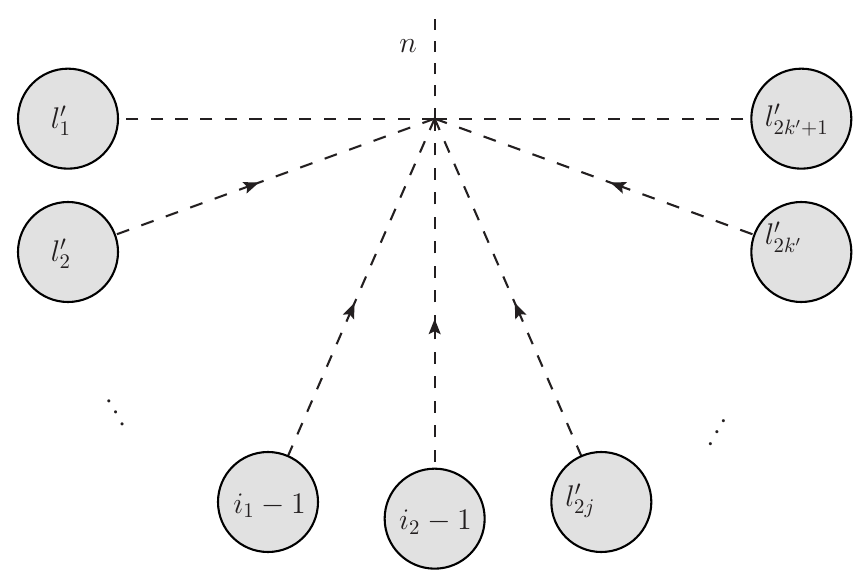}
\caption{\label{figtsc2p2} A diagram when two soft legs $i_1$ and $i_2$ split a $J$ into three parts, with odd $i_1$ and $i_2$. }
\end{figure}

Case 1-ii) is similar to case 1-i). Here we have both $i_1$ and $i_2$ to be odd, and neither of them is adjacent to $n$. We have the same 4 scenarios as in case 1-i), while we need to slightly modify the diagrams. This time, when a soft leg is attached to a non-trivial $J$ of label $l_m$, $m$ need to be even instead of odd to generate a $\ordr (\tau)$ contribution in the triple soft limit. Then we have diagrams shown in Figs. (\ref{figtsc2p1}) and (\ref{figtsc2p2}). The momentum factor will be $2 p_n \cdot (q_{i_1-1} + q_{i_2-1} + \sum_j q_{l'_{2j}})$. In the end, the first and second scenarios give
\bea
\tau \sum_{k=1}^{n/2-3} \frac{(-1)^k}{f^{2k+4}} \sum_{\{l_m \}} 2 p_{n} \cdot \left( q_{i_1-1} + q_{i_2-1}+  \sum_{j=1}^k q_{l'_{2j}}  \right)  \prod_{m=1}^{2k+3} J (l_{m-1}+1, \cdots , l_m) +\ordr (\tau^2).
\eea
The third scenario gives
\bea
&&2\tau \sum_{k=2}^{n/2-2} \frac{(-1)^k}{f^{2k+2}} \sum_{\{l_m \}} 2 p_{n} \cdot   \left( q_{i_1-1} + q_{i_2-1}+  \sum_{j=1}^{k-1} q_{l'_{2j}}  \right)  \prod_{m=1}^{2k+1} J (l_{m-1}+1, \cdots , l_m) +\ordr (\tau^2)\non\\
&=&-2\tau \sum_{k=1}^{n/2-3} \frac{(-1)^k}{f^{2k+4}} \sum_{\{l_m \}} 2 p_{n} \cdot   \left( q_{i_1-1} + q_{i_2-1}+  \sum_{j=1}^k q_{l'_{2j}}  \right) \prod_{m=1}^{2k+3} J (l_{m-1}+1, \cdots , l_m) +\ordr (\tau^2).\non\\
\eea
Note that in the first line the $k=1$ case does not exist, as neither of leg $i_1$ nor leg $i_2$ is adjacent to leg $n$. Similarly, the fourth scenario is
\bea
&&\tau \sum_{k=3}^{n/2-1} \frac{(-1)^k}{f^{2k}} \sum_{\{l_m \}} 2 p_{n} \cdot   \left( q_{i_1-1} + q_{i_2-1}+  \sum_{j=1}^{k-2} q_{l'_{2j}}  \right) \prod_{m=1}^{2k-1} J (l_{m-1}+1, \cdots , l_m) +\ordr (\tau^2)\non\\
&=&\tau \sum_{k=1}^{n/2-3} \frac{(-1)^k}{f^{2k+4}} \sum_{\{l_m \}} 2 p_{n} \cdot   \left( q_{i_1-1} + q_{i_2-1}+  \sum_{j=1}^k q_{l'_{2j}}  \right)  \prod_{m=1}^{2k+3} J (l_{m-1}+1, \cdots , l_m) +\ordr (\tau^2).\qquad
\eea
Again, all the $\ordr (\tau )$ terms cancel out, so that $M_{n}^{\text{\nlsm}} (\mathbb{I}_{n}) = \ordr (\tau^2)$.

\subsection{Case {2}}

Here we have $i_1 = 1$ and $i_2>2$, and $i_2$ is even. From Eq. (\ref{eq:Jsinglesoft}) we know that if leg $i_1$ is attached to a non-trivial $J$, we get zero. Therefore, leg $i_1$ must be attached to the same vertex as leg $n$. Then we only have two scenarios, where leg $i_2$ is either attached to a non-trivial $J$ or a trivial $J$.

\begin{figure}[htbp]
\centering
\includegraphics[width=0.5\textwidth]{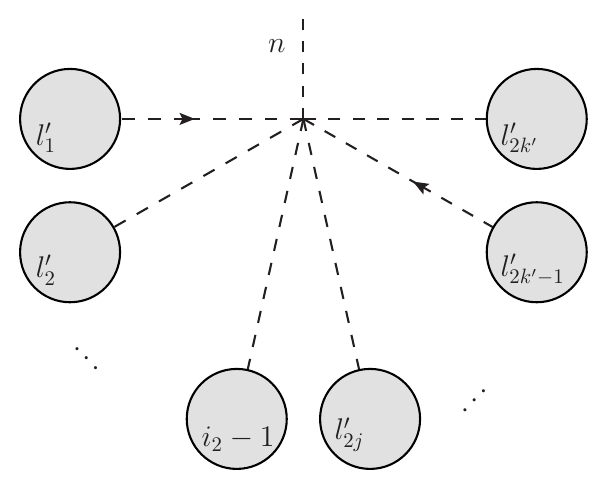}
\caption{\label{figtsc3p1} A diagram when $i_2$ split a $J$ in half, with $i_1=1$ and $i_2$ even. }
\end{figure}

\begin{figure}[htbp]
\centering
\subfloat[]{
\centering
\includegraphics[height=1.2in]{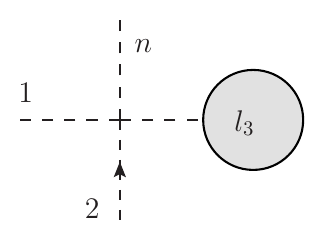}
\label{figtsc3s1}
}
\qquad
\subfloat[]{
\centering
\includegraphics[height=1.2in]{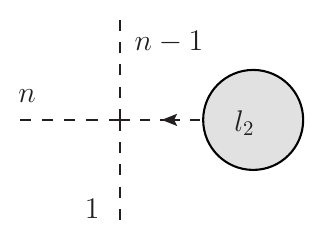}
\label{figtsc3s2}
}
\caption{\label{figtsc3s} The two ``pole diagrams'' in case {2}, which need to be treated carefully.}
\end{figure}

When $i_2$ is attached to a non-trivial $J$, we have the diagram shown in Fig. \ref{figtsc3p1}, where we have $2k+1$ $J$'s so that $k'=k$. The $k=1$ case needs to be treated separately, where we can have a ``pole diagram'' as shown in Fig. \ref{figtsc3s1}. Naively the 4-point vertex is at $\ordr (\tau)$ and leg $i_2$ just split the $J$ labeled by $l_3$. However, the off-shell leg in $J$ has momentum $p_I = p_2+ \tau (p_1 + p_n)$, and as $J$ always contains the propagator $i/p_I^2$, we now have an additional pole in $\tau$ from the propagator. The consequence is that apart from the ``splitting'' term given by Eq. (\ref{eq:Jsinglesoft}), the $J$ labeled by $l_3$ generate an additional term at $\ordr (1)$:
\bea
&&  \frac{ i}{p_2 \cdot (p_1 + p_n)} \sum_{\substack{j=2\\ j \ne i_2, i_2 \pm 1}}^{n-1} s_{i_2,j}\non\\
 &&\times M_{n-3}^{\text{\nlsm}\oplus \phi^3}( i_2 + 1, \cdots, n-1,2,\cdots,i_2 - 1,|i_2+1,i_2-1,j) + \ordr (\tau^2),\qquad
\eea
which is the propagator of the off-shell leg times the single soft limit of the on-shell amplitude $M_{n-2}^{\text{\nlsm}} (2,3,\cdots,n-1)$ with $i_2$ soft. Clearly, this term is $\ordr (\tau)$ when there is no pole in the propagator $i/p_I^2$, so we do not consider it elsewhere. Therefore, we have an additional contribution in the triple soft limit,
\bea
&& \frac{\tau }{f^{2}}  \frac{ p_{n} \cdot    p_2}{p_2 \cdot (p_1 + p_n)} \sum_{\substack{j=2\\ j \ne i_2, i_2 \pm 1}}^{n-1} s_{i_2,j}\non\\
 &&\times M_{n-3}^{\text{\nlsm}\oplus \phi^3}( i_2 + 1, \cdots, n-1,2,\cdots,i_2 - 1,|i_2+1,i_2-1,j) + \ordr (\tau^2),\qquad\label{eqtsc3s1}
\eea
Similarly, the other pole diagram Fig. \ref{figtsc3s2} gives an additional contribution
\bea
&&-\tau  \frac{\tau }{f^{2}}  \frac{ p_{n} \cdot    p_{n-1}}{p_{n-1} \cdot (p_1 + p_n)} \sum_{\substack{j=2\\ j \ne i_2, i_2 \pm 1}}^{n-1} s_{i_2,j}\non\\
 &&\times M_{n-3}^{\text{\nlsm}\oplus \phi^3}(i_2 + 1, \cdots, n-1,2,\cdots,i_2 - 1,|i_2+1,i_2-1,j) + \ordr (\tau^2).\qquad\label{eqtsc3s2}
\eea
On the other hand, the normal terms where leg $i_2$ just ``split'' diagrams give
\bea
 \tau \sum_{k=1}^{n/2-2} \frac{(-1)^k}{f^{2k+2}} \sum_{\{l_m \}} 2 p_{n} \cdot   \sum_{j=1}^k q_{l'_{2j-1}}  \prod_{m=1}^{2k+1} J (l_{m-1}+1, \cdots , l_m) +\ordr (\tau^2),
\eea
where $\{ l_m \}$ is a  splitting of the ordered set $\{2,3,\cdots,i_2-1, i_2+1, \cdots n-1\}$ into $2k+3$ non-empty ordered subsets $\{l_{m-1}+1,l_{m-1}+2,\cdots,l_m\}$  (here $l_0 = 1$ and $l_{2k+1} = n-1$), with $i_2-1 \in \{l_m\}$, and the other elements are denoted in order by $l'_m$. Note the non-existence of diagrams with $k=n/2-1$.

When $i_2$ is attached to a trivial $J$, the diagram will be the same as in Fig. \ref{figtsc3p1}, except that we only have $2k-1$ $J$'s and $k' = k-1$. This gives
\bea
&& \tau \sum_{k=2}^{n/2-1} \frac{(-1)^k}{f^{2k}} \sum_{\{l_m \}} 2 p_{n} \cdot   \sum_{j=1}^{k-1} q_{l'_{2j-1}}  \prod_{m=1}^{2k-1} J (l_{m-1}+1, \cdots , l_m) +\ordr (\tau^2)\non\\
&=& -\tau \sum_{k=1}^{n/2-2} \frac{(-1)^k}{f^{2k+2}} \sum_{\{l_m \}} 2 p_{n} \cdot   \sum_{j=1}^k q_{l'_{2j-1}}  \prod_{m=1}^{2k+1} J (l_{m-1}+1, \cdots , l_m) +\ordr (\tau^2).\qquad\label{egtsc3p2}
\eea
Note the non-existence of diagrams with $k=1$ in the first line of Eq. (\ref{egtsc3p2}), as $i_2 > 2$. In the end, only the terms in  Eqs. (\ref{eqtsc3s1}) and (\ref{eqtsc3s2}) survive, and we arrive at Eq. (\ref{eqtsc2}).

\subsection{Case {3}}

\begin{figure}[htbp]
\centering
\subfloat[]{
\centering
\includegraphics[height=1.2in]{tsdac3s1-eps-converted-to.pdf}
\label{figtsc4s1}
}
\qquad
\subfloat[]{
\centering
\includegraphics[height=1.2in]{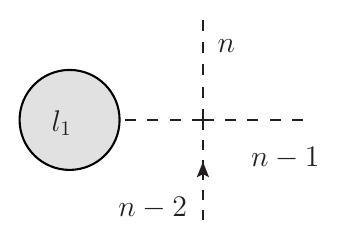}
\label{figtsc4s2}
}
\caption{\label{figtsc4s} The two ``pole diagrams'' in case {3}.}
\end{figure}

When $i_1=1$ and $i_2 = n-1$, to contribute to $\ordr (\tau)$ terms in the triple soft limit, neither of the legs $1$ and $n-1$ can be attached to a non-trivial $J$, except for the ``pole diagrams'' shown in Fig. \ref{figtsc4s}. These pole diagrams give
\bea
&&  \frac{\tau }{f^{2}}\sum_{j=3}^{n-3}  \left[ \frac{ p_{n} \cdot    p_2}{p_2 \cdot (p_1 + p_n)} s_{n-1,j}+ \frac{ p_{n} \cdot    p_{n-2}}{p_{n-2} \cdot (p_{n-1} + p_n)} s_{1,j} \right]\non\\
&&\times  M_{n-3}^{\text{\nlsm}\oplus \phi^3}(2,3,\cdots, n-2|2,n-2,j) + \ordr (\tau^2).
\eea
On the other hand, both leg $1$ and $n-1$ can be attached to the same vertex as leg $n$, which gives
\bea
&& \tau \sum_{k=1}^{n/2-1} \frac{(-1)^k}{f^{2k}} \sum_{\{l_m \}} 2 p_{n} \cdot   \sum_{j=1}^k q_{l_{2j-1}}  \prod_{m=1}^{2k-1} J (l_{m-1}+1, \cdots , l_m)+ \ordr(\tau^2)\non\\
 &=&\frac{\tau}{f^2} \sum_{k=1}^{n/2-2} \frac{(-1)^k}{f^{2k}} \sum_{\{l_m \}} 2 p_{n} \cdot   \sum_{j=1}^k q_{l_{2j}}  \prod_{m=1}^{2k+1} J (l_{m-1}+1, \cdots , l_m)\non\\
&&- \frac{\tau }{f^{2}}  \frac{ 2p_{n} \cdot   (p_1 + p_{n-1}) }{ (p_1+ p_{n-1}  + p_n)^2} \sum_{j=3}^{n-3} \left( s_{n-1,j} + s_{n,j} + s_{1,j} \right) \non\\
 &&\times M_{n-3}^{\text{\nlsm}\oplus \phi^3}(2,3,\cdots, n-1|2,n-2,j) + \ordr (\tau^2)\non\\
 &=&\frac{\tau}{f^2} \sum_{j=3}^{n-3}  \frac{2 ( p_1 \cdot p_{n-1}) s_{n,j} -  2p_{n} \cdot   (p_1 + p_{n-1})\left( s_{n-1,j} + s_{1,j} \right) }{ (p_1+ p_{n-1}  + p_n)^2} \non\\
 &&\times M_{n-3}^{\text{\nlsm}\oplus \phi^3}(2,3,\cdots,n-2|2,n-2,j) + \ordr(\tau^2),
\eea
where in the first and second  line $\{ l_m \}$ is a  splitting of the ordered set $\{2,3,\cdots,n-3,n-2\}$ into $2k- 1$ and $2k+1$ non-empty ordered subsets $\{l_{m-1}+1,l_{m-1}+2,\cdots,l_m\}$, respectively. The first equality comes from total momentum conservation, as well as singling out the $k=1$ case in the first line. Summing over all contributions and we get Eq. \ref{eqtsc3}.

\bibliographystyle{JHEP}
\bibliography{references_amp}

\end{document}